\documentclass[11pt,a4paper]{article}

\usepackage{geometry}
\usepackage{graphicx,stfloats}
\usepackage{color}

\usepackage{ulem}
\usepackage{amssymb}
\usepackage{amsmath}
\usepackage{url}
\usepackage{lipsum}

\newcommand{\D}{\mathrm{d}}

\newcommand{\Ka}{\mathrm{Ka}}

\title{An {\it a priori} analysis of a DNS database of turbulent lean premixed methane flames for LES with finite-rate chemistry}

\author{A. J. Aspden$^1$, N. Zettervall$^2$ and C. Fureby$^2$}

\date{\small{$^1$School of Engineering, Newcastle University, Stephenson Building,\\
Claremont Road, Newcastle-Upon-Tyne, NE1 7RU, UK\vspace{2mm}\\
$^2$Defence Security Systems Technology, The Swedish Defence Research Agency,\\
FOI, SE 147 25 Tumba, Stockholm, Sweden}\vspace{4mm}\\
2nd May 2018}

\begin{document}

\maketitle

\begin{abstract}
An {\it a priori} analysis of a DNS database of turbulent lean premixed methane flames is 
presented considering the relative effects of turbulence and LES filtering, 
along with a potential modelling approach for LES with finite-rate chemistry.  
The leading-order effect was found to be due to the filter operation; 
flame response to turbulence was a secondary effect, and manifested primarily as an increase in 
standard deviations about conditional means.  It was found that the radicals O, H and OH 
were less impacted by the filter than other high-temperature radicals, which were significantly 
reduced in magnitude by the filter.  By considering reaction path diagrams, key reactions
have been identified that are responsible for disparities between the desired filtered reaction
rates and the reaction rates evaluated using quantities available in LES calculations (i.e.\ the 
filtered species and temperature).  More specifically, the hydrogen abstraction reactions that take 
CH$_4$ to CH$_3$ (by O, H and OH) were found to have particularly enhanced reaction rates, and 
dominate the reaction path.  Under the conditions presented, reaction paths were found to be 
independent of turbulence intensity.  In general, the filtered reaction rates from the DNS were 
found to align more closely with the filtered laminar profile than the reaction rate of filtered 
species and temperature, (but disparities were found to decrease with increasing Karlovitz 
number).  A simple model for scaling reaction rates is considered based on filtered laminar 
flame profiles, and the resulting reaction paths demonstrate proof-of-concept of a simple
approach for formulating a reaction rate model for LES with finite-rate chemistry.
\end{abstract}

\section{Introduction}
\label{Introduction}

Predictive modelling of turbulent combustion incorporating finite-rate kinetics is becoming 
increasingly important for the development of fuel-flexible combustion devices 
with low emissions and high-speed propulsion systems.
Large Eddy Simulation (LES) is a promising approach requiring closure models describing subfilter 
transport as for non-reactive flows \cite{sagaut2006large}, but also for filtered reaction rates,
\cite{poinsot2005theoretical,janicka2005large,echekki2009multiscale,menon2010computational}.
Mixing and chemical reactions usually occur together on scales smaller than convection, requiring 
different modelling approaches, \cite{echekki2009multiscale,menon2010computational}. 
These include flamelet models, e.g.\ \cite{hawkes2000flame}, finite-rate chemistry models such as 
thickened flame models, \cite{colin2000thickened}, 
localized time-scale models, \cite{fureby2009large,giacomazzi2000fractal,sabelnikov2013combustion}, 
approximate deconvolution models, \cite{mathew2002large}, 
presumed Probability Density Function (PDF) models, \cite{gerlinger2003investigation}, 
transported PDF models, \cite{bulat2013large,kim2014effects}, 
Conditional Moment Closure (CMC) models, \cite{navarro2005conditional}, 
and Linear Eddy Models (LEM), \cite{menon2011linear}. 

Assessment of flamelet and finite-rate chemistry models,
e.g.~\cite{hernandez2011laboratory,ma2014posteriori,fureby2017comparative,fedina2017assessment},
have found satisfactory overall agreement for most models, but with finite-rate chemistry models 
performing somewhat better than the flamelet models.
In finite-rate chemistry LES, 
the combustion chemistry is incorporated by solving filtered transport equations for the species, 
with the filtered reaction rates being computed either explicitly using 
filtered Arrhenius reaction rate expressions or tabulated reaction rates, \cite{bulat2015reacting}.  
An issue with finite-rate chemistry models is the dependence on the underlying reaction 
mechanism, \cite{bulat2015reacting,zettervall2017large}.
From \cite{bulat2015reacting,zettervall2017large,fiorina2005premixed} and other similar studies, it 
has been observed that skeletal reaction mechanisms can successfully be used in finite-rate chemistry 
LES.

The influence of the filtering is handled through different types of mathematical or 
phenomenological models, e.g.\ \cite{colin2000thickened,fureby2009large,giacomazzi2000fractal,sabelnikov2013combustion,mathew2002large,gerlinger2003investigation,bulat2013large,kim2014effects,navarro2005conditional,menon2011linear}, in conjunction with the reaction rates. 
The filtering operation has been examined previously for non-reacting LES 
(e.g.\ \cite{liu1994properties}), and for combustion LES \cite{LapointeCnF17} 
targetting tabulated chemistry based on presumed PDFs. 
Here, Direct Numerical Simulation (DNS) results from lean premixed methane-air flames 
will be used to examine the influence of the underlying turbulent flow on the filtered reaction
rates, and hence the modelling requirements for finite-rate chemistry LES.

\section{LES of Turbulent Combustion}

LES equations of motion are derived from conservation of mass, momentum and energy by applying
a low-pass filter.  Physical processes on scales larger than the filter width, $\Delta$, are 
resolved, whereas physics occurring on scales smaller than $\Delta$ require subfilter models. 
The full equation set can be found elsewhere (e.g.\ \cite{echekki2009multiscale}), but the focus
here is on the reaction terms appearing in the conservation of mass for species $i$,
\begin{align}
\frac{\partial}{\partial t}\left(\bar{\rho}\tilde{Y}_i\right)
&+\nabla\cdot\left(\bar{\rho}\tilde{Y}_i\tilde{\boldsymbol{v}}\right)
=\nabla\cdot\left(D_i\nabla\tilde{Y}_i-\boldsymbol{b}_i\right)
+\bar{\dot{\omega}}_i,
\label{eq:LES2}
\end{align}
in which, $\bar{\rho}$, $\tilde{\boldsymbol{v}}$, and $\tilde{Y}_i$ 
are the (Favre) filtered density, velocity, and species mass fractions, respectively, 
$D_i$ is the Fickian diffusion coefficient for species $i$,
and the subfilter turbulent mixing is hidden in the diffusive flux term $\boldsymbol{b}_i$.
The filtered reaction term $\bar{\dot{\omega}}_i$ is the focus of the present work, 
in particular the non-linear response to the LES filtering operation,
which can be written out to emphasize the dependencies on all dependent variables,
\begin{equation}
\bar{\dot{\omega}}_i=M_i\sum_{j=1}^M (P_{ij}^{\prime\prime}-P_{ij}^{\prime}) \bar{Q}_j,
\label{eq:reac}
\end{equation}
where $\bar{Q}_j$ are the filtered progress rates of reaction $j$,
\begin{equation}
\bar{Q}_j=\overline{\left[k_{f,j}\prod_{k=1}^N\left(\frac{\rho Y_k}{M_k}\right)^{P_{kj}^{\prime}}
                         -k_{b,j}\prod_{k=1}^N\left(\frac{\rho Y_k}{M_k}\right)^{P_{kj}^{\prime\prime}}\right]},
\label{eq:reacFilt}
\end{equation}
where $k_{f,j}$ and $k_{f,j}$ are the forward and backward rates of reaction $j$, respectively.
Taylor series expansions of (\ref{eq:reacFilt}) have been discussed (from a RANS point of view) in 
\cite{poinsot2005theoretical}, 
and this is not considered a useful approach for increasing the understanding of these 
terms or for the development of improved models due to the inherent non-linearites. Alternatively, 
by multiplying and dividing each of the reaction rates in (\ref{eq:reac}) by the filtered 
reaction rates we have,
\begin{equation}
\bar{\dot{\omega}}_i=M_i\sum_{j=1}^M(P_{ij}^{\prime\prime}-P_{ij}^{\prime})\Omega_jQ_j(\tilde{\boldsymbol{Y}},\bar{T}),
\end{equation}
in which 
$\Omega_j=\bar{Q}_j/Q_j(\tilde{\boldsymbol{Y}},\bar{T})$
denote the correlations between the filtered reaction 
rates and the reaction rates evaluated by the filtered quantities accessible in LES. 
The correlations $\Omega_j$ then constitute a model approach for LES with finite-rate chemistry.
The premise of the present work is to evaluate factors affecting these 
correlation terms for a range of lean premixed flames at different $\Ka$ to investigate the 
influence of the turbulence on the combustion chemistry, and thus also obtain information about 
the subfilter modelling requirements for finite-rate chemistry LES.

\section{DNS of Turbulent Premixed Methane Flames}

\begin{figure}
\centering
\includegraphics[width=88mm]{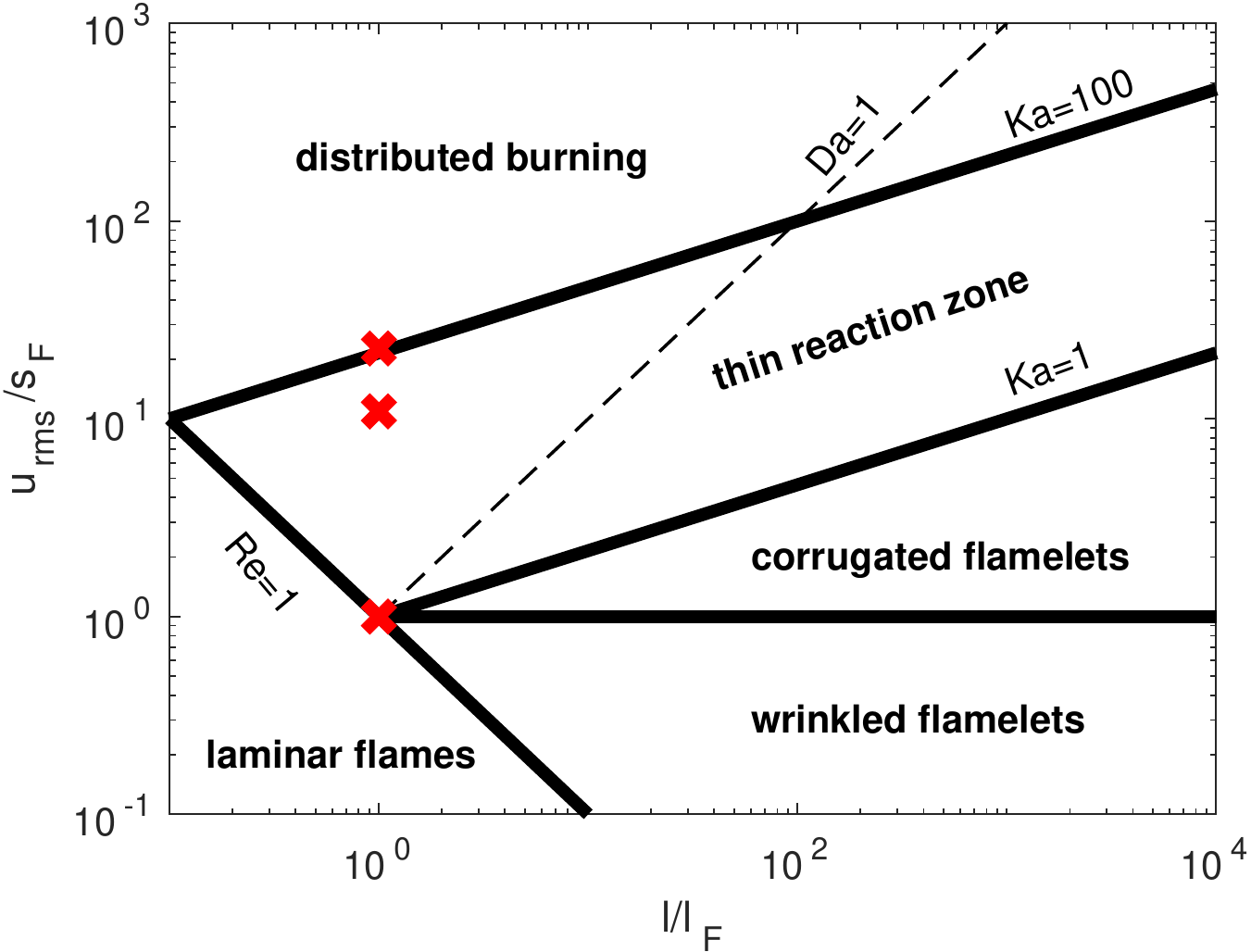}
\caption{Regime diagram showing the conditions analysed.}
\label{fig:regime}
\end{figure}

The simulation database that will be used for the present study consists of a series of DNS with 
detailed chemistry of statistically-stationary statistically-planar 
turbulent premixed methane flames in maintained homogeneous isotropic turbulence,
\cite{AspdenCNF16,AspdenDodecane17,AspdenHiKa18}.
The simulations were run using the well-established low Mach number combustion solver developed at 
the Center for Computational Sciences and Engineering at the Lawrence Berkeley National Laboratory.  
The details of the numerical method can be found in \cite{DayBell2000} and \cite{Nonaka2012}.  
The methodology treats the fluid as a mixture of perfect gases, using a mixture-averaged model 
for diffusive transport, ignoring Dufour and Soret effects.
A long-wavelength forcing term designed to establish and maintain turbulence with the desired 
properties \cite{Aspden08b}.  
The chemical kinetics and transport were modelled using the GRIMech 3.0 without emissions chemistry 
\cite{FrenklachWang1995}, resulting in 35 species with 217 elementary reactions.  
The simulations were conducted at $\Lambda=l/l_F=1$, 
as part of the study reported in \cite{AspdenDodecane17}, matching the Karlovitz numbers of the 
$\Lambda=4$ calculations reported in \cite{AspdenCNF16} 
($\Ka=(u^3l_F)/(s_F^3l)=1$ and 36), along with a higher 
Karlovitz number case from \cite{AspdenHiKa18} ($\Ka=108$) looking at more turbulent conditions.
The conditions are shown on the regime diagram in figure~\ref{fig:regime},
and span the conventionally-defined thin reaction zone.
As with all DNS studies, the integral length scale has been sacrificed to resolve the flame
adequately, but is sufficient for studying small-scale turbulence-chemistry interaction, and is 
representative of high intensity turbulence that would reach these scales through the energy 
cascade from larger integral lengths at the same Karlovitz numbers.

\section{Analysis of the DNS Data}

The DNS data were filtered using a simple top-hat box filter with a width approximately equal to
one flame thermal thickness; at approximately 660 microns, this filter width is typical 
for combustion LES (e.g.\ \cite{fureby2017comparative}), 
and much larger than the scales over which the reactions take place.
A quantity $q$ filtered in this way will be denoted as $\bar{q}$,
with Favre filtered quantities denoted $\tilde{q}=\overline{\rho q}/\bar{\rho}$.
Comparisons are made between the reaction rates $Q_j$, the filtered reaction rates 
$\bar{Q}_j$ and the reaction rates evaluated using filtered species and temperature
$Q_j(\tilde{\boldsymbol{Y}},\bar{T})$.  Each of these three kinds of reaction rates was evaluated
for a laminar flame profile (i.e.~a steady unstrained one-dimensional flame) and for the DNS data, 
from which a temporally-averaged  mean and standard deviation was evaluated conditioning on 
temperature.

\begin{figure}
\centering
\makebox[38mm][c]{\small $\Ka=1$}\makebox[38mm][c]{\small $\Ka=108$}\\
\makebox[19mm][c]{\small unfiltered}\makebox[19mm][c]{\small filtered}
\makebox[19mm][c]{\small unfiltered}\makebox[19mm][c]{\small filtered}\\
\includegraphics[width=38mm]{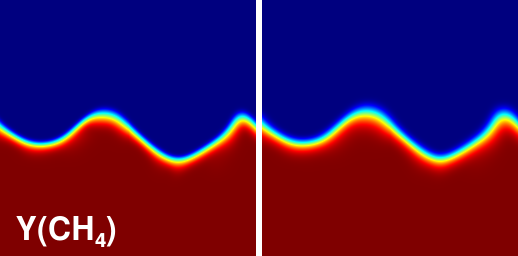}     % ch4
\includegraphics[width=38mm]{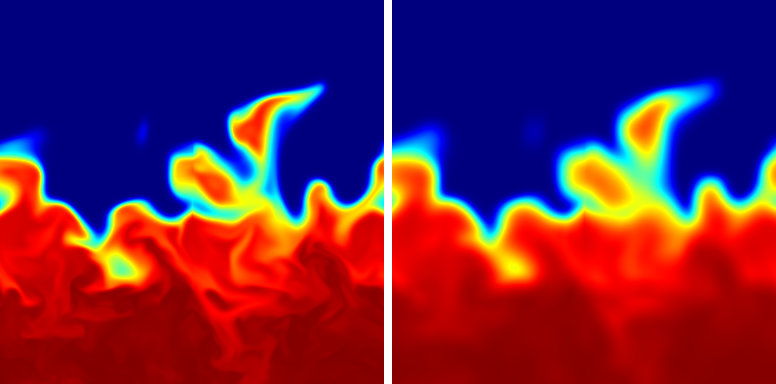} \\ % ch4
\includegraphics[width=38mm]{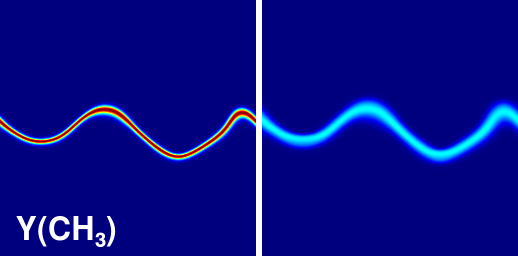}     % ch3
\includegraphics[width=38mm]{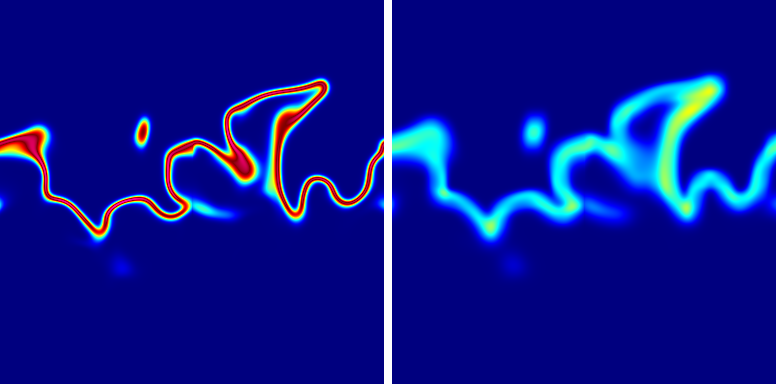} \\ % ch3
\includegraphics[width=38mm]{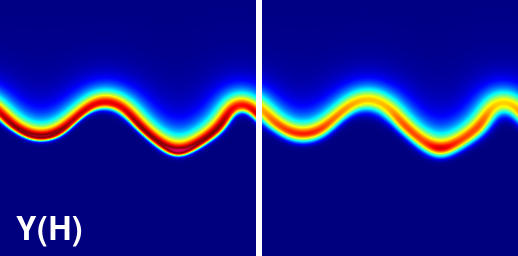}     % h
\includegraphics[width=38mm]{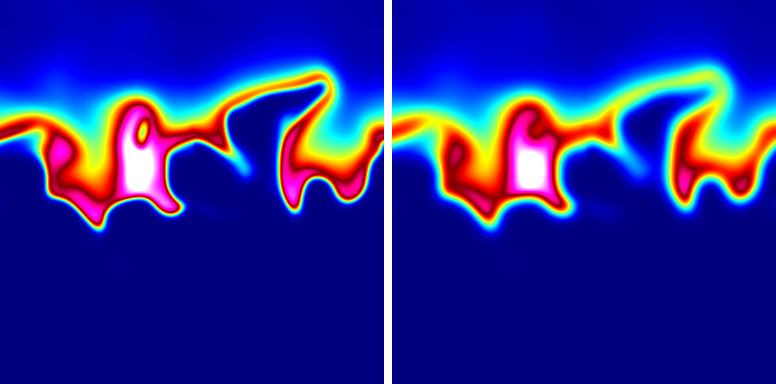} \\ % h
\includegraphics[width=38mm]{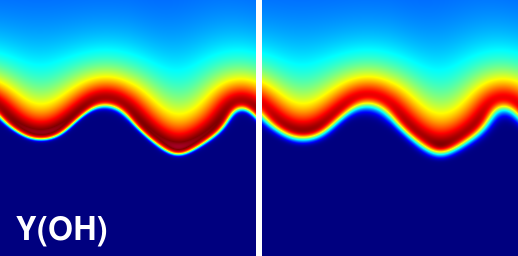}     % oh
\includegraphics[width=38mm]{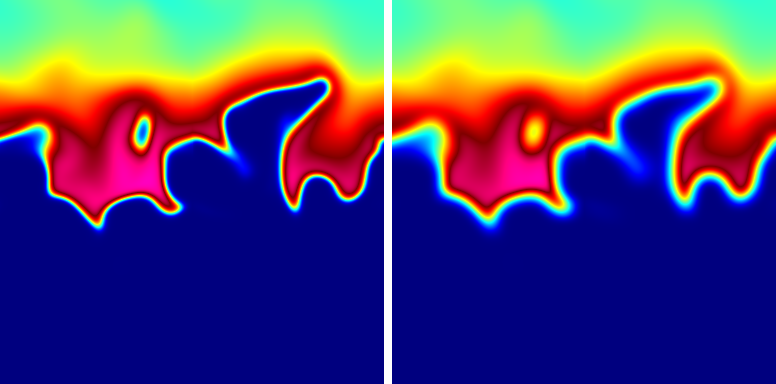} \\ % oh
\includegraphics[width=38mm]{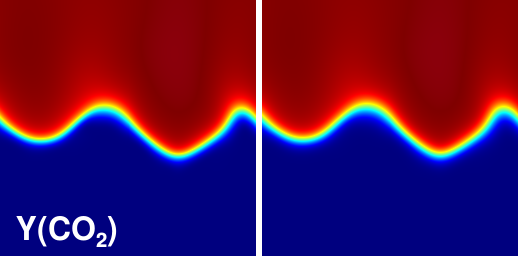}     % co2
\includegraphics[width=38mm]{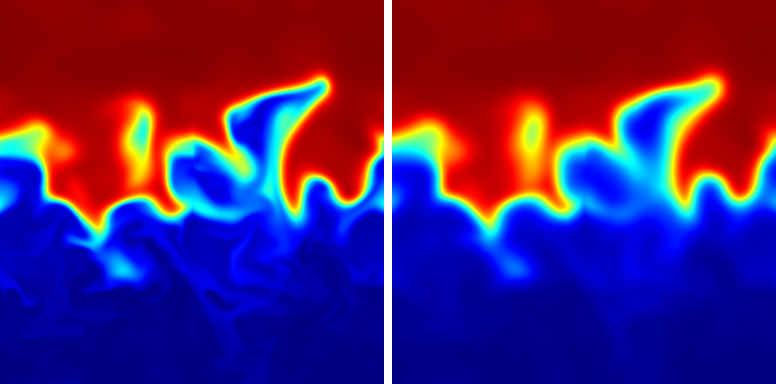} \\ % co2
\includegraphics[width=60mm]{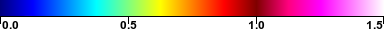}
\caption{Two-dimensional slices comparing the effect of the filter at $\Ka=1$ and 36;
all fields are normalised by corresponding laminar values, and periodicity has been 
exploited to join $x$ and $y$ slices together to show more flame surface.  Each square panel
is 20 flame thickness in size.}
\label{fig:slices}
\end{figure}

The effect of turbulence and the filtering procedure on some representative species is presented 
in figure~\ref{fig:slices} as two-dimensional slices; all species at all $\Ka$ are presented in 
the supplementary material.  As expected, the turbulence has little effect at $\Ka=1$ other than 
producing large-scale
wrinkling.  At $\Ka=108$, turbulence has a significant effect on the flame, especially on the
preheat region, which is substantially thickened (e.g.\ CH$_4$).  
Turbulence also increases the occurrence of highly-curved regions, which leads to variation along 
the flame surface and a slightly enhanced radical pool (e.g.\ H).  
Further details of flame response to turbulence in these cases can be found in 
\cite{AspdenCNF16,AspdenDodecane17,AspdenHiKa18}.
Naturally, the filter smooths out monotonic fields like fuel mass fraction, but has a more
significant effect particularly on the thickness and magnitude of narrow fields such as CH$_3$.
At high $\Ka$, the radicals H and OH appear less impacted by the filter, and present values 
in excess of the laminar values (as seen by the magenta and white regions).

\begin{figure}
\centering
\includegraphics[width=48mm]{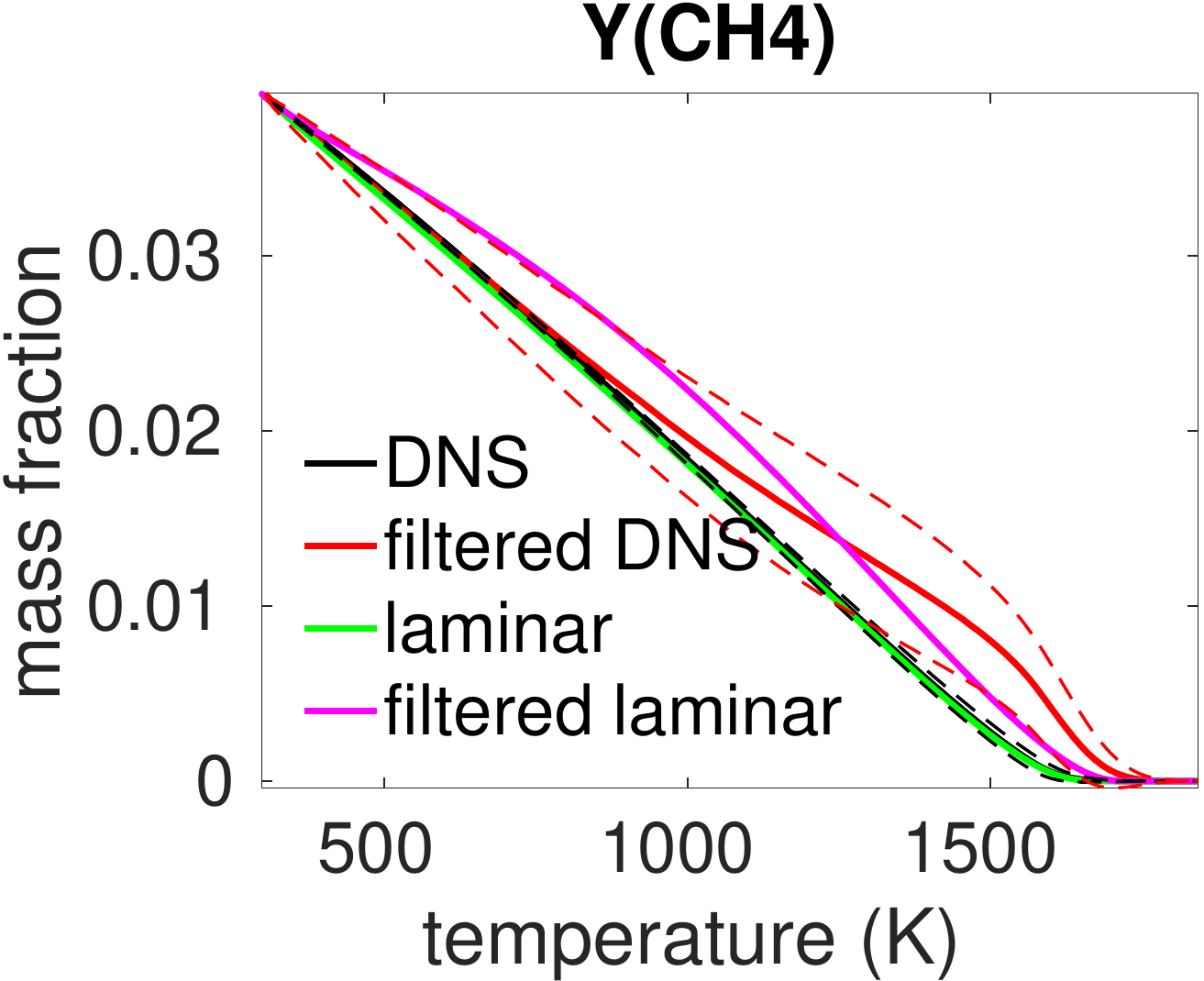}
\includegraphics[width=48mm]{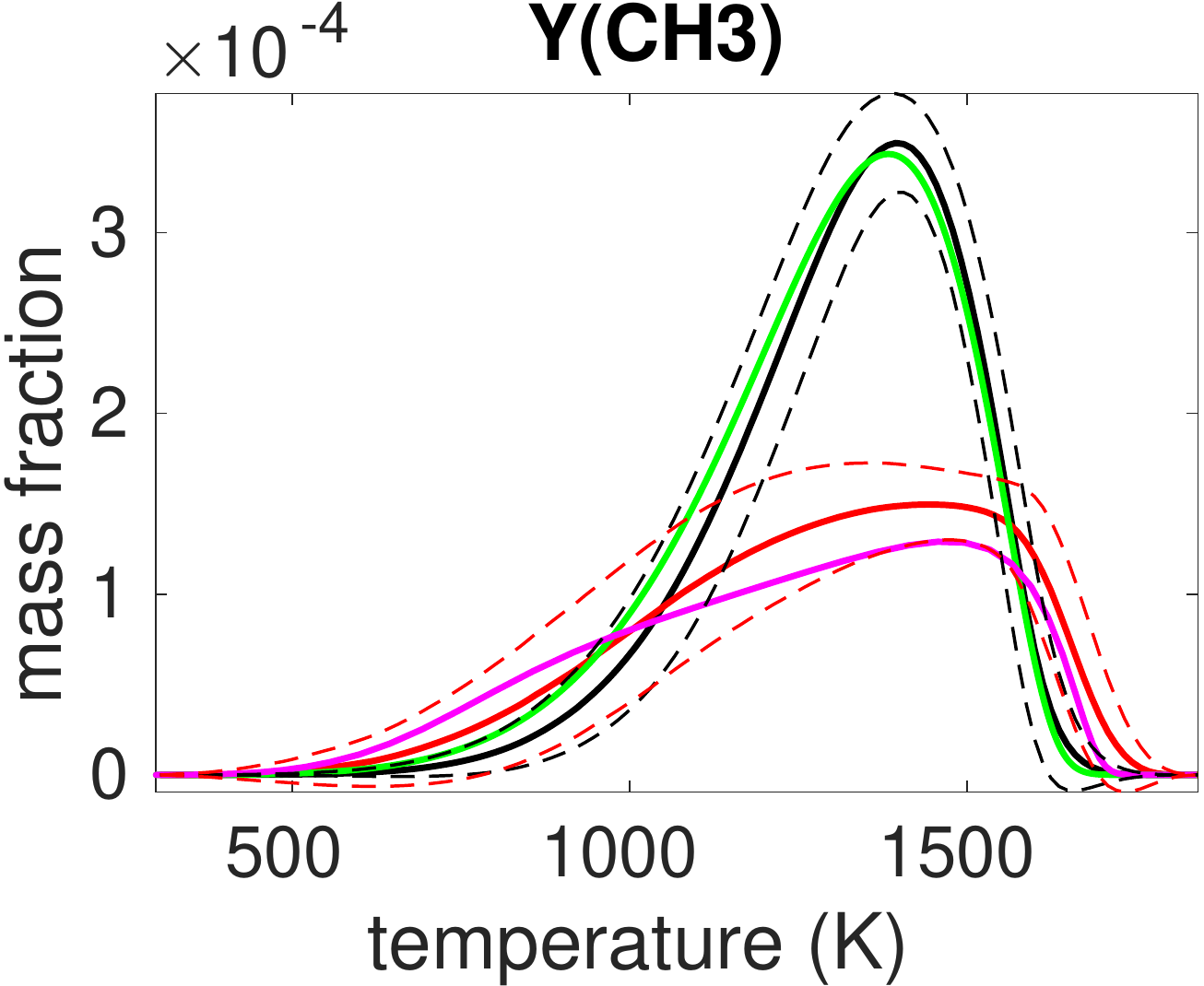}\vspace{2mm}\\
\includegraphics[width=48mm]{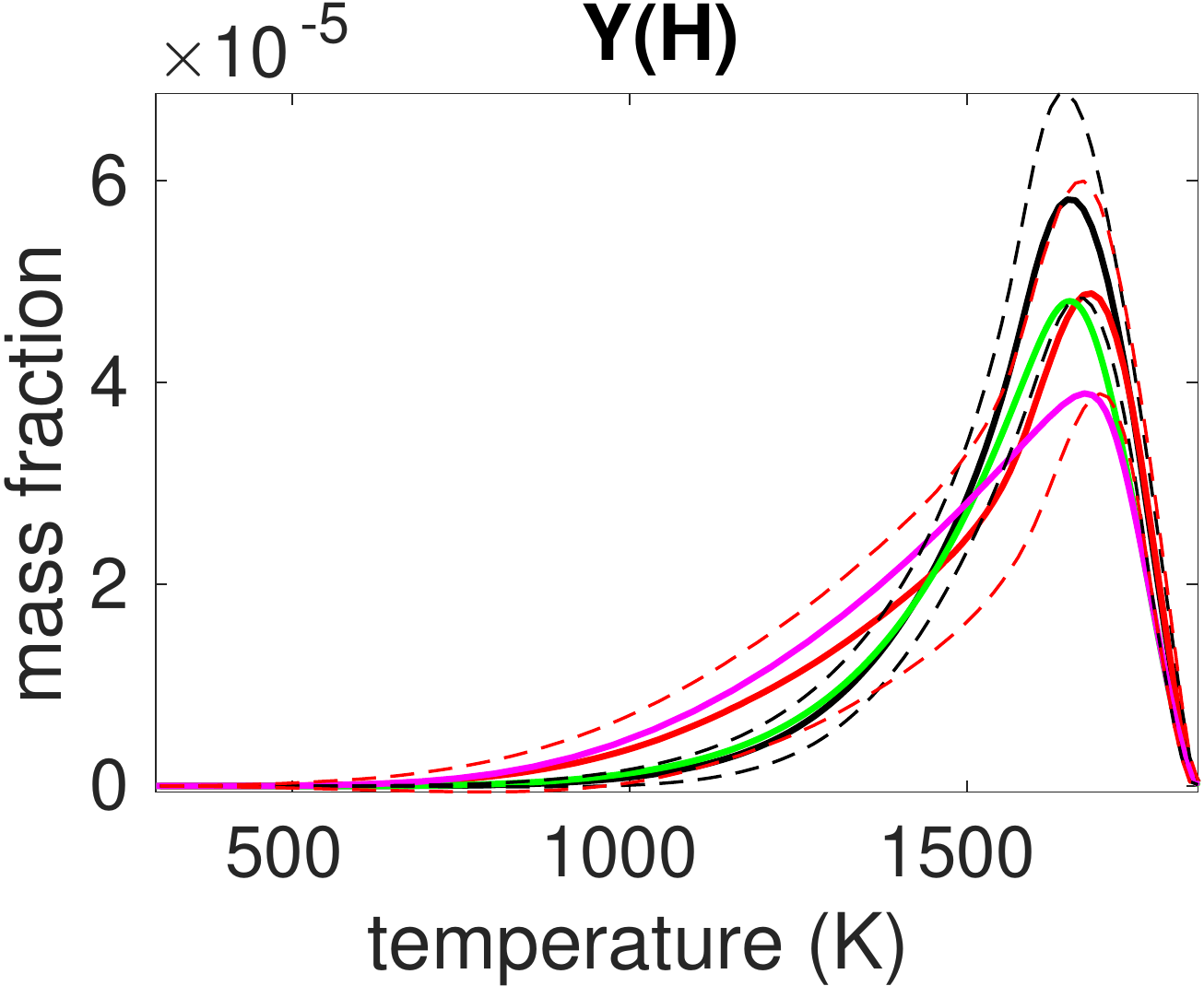}
\includegraphics[width=48mm]{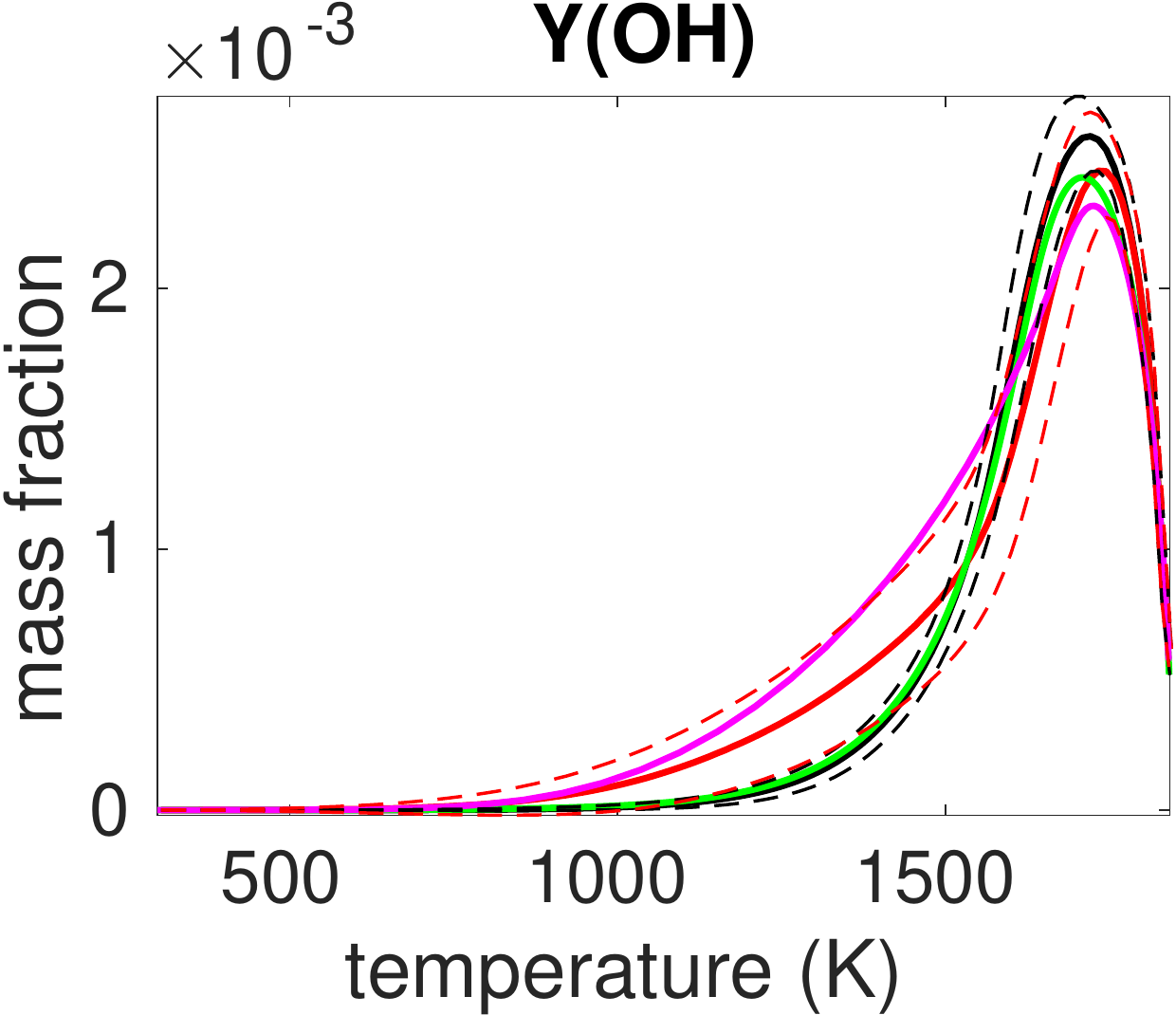}
\caption{Conditional means of unfiltered and filtered species mass fractions for $\Ka=108$.
One standard deviation about the mean are shown by dashed lines of corresponding colour.
Note that the filtered profiles are conditioned on filtered temperature.}
\label{fig:species}
\end{figure}

The response of species distribution to turbulent mixing was classified in \cite{AspdenCNF16}, and 
the response to the filtering operation has been found to be consistent with this classification; 
all species profiles for all $\Ka$ are presented by classification in the supplementary material,
but the profiles for CH$_4$, CH$_3$, H and OH at $\Ka=108$ are presented in figure~\ref{fig:species}.
At low $\Ka$ (again see supplementary material), the conditional means align closely with the 
laminar profiles, and the standard deviations are small.  The main response to increasing the 
Karlovitz number, is an increase in the standard deviations for intermediate species
(see dashed lines in figure~\ref{fig:species}).  The filtered profiles are more 
interesting, and depend on temperature, Karlovitz number, and species type.  Reactants and 
products (see supplementary data) present filtered profiles that, at low-to-moderate temperatures,
align with the unfiltered turbulent profile (it is likely that both actually align with the unity 
Lewis number profile); at higher temperatures, however, there is a deviation from the other profiles 
(specifically, an increase in fuel mass fraction and a decrease in water mass fraction), the 
explanation for which is currently unclear.  
The alignment at low temperatures is attributed to penetration 
of turbulence into the preheat region, broadening the profiles in physical space.
The conditional means of the radicals O, H and OH increase, whereas almost all of the other 
carbonated radicals (represented here by CH$_3$) remain close to the laminar profile.  A significant
difference is observed in response to the filter operation; a substantial decrease in the peak value
is observed in the short-lived high-temperature radicals, but is much less pronounced for O, H and 
OH, which present increased profiles at lower temperatures.

\begin{figure}
\centering
\includegraphics[trim=140 175 150 125,clip,width=48mm]{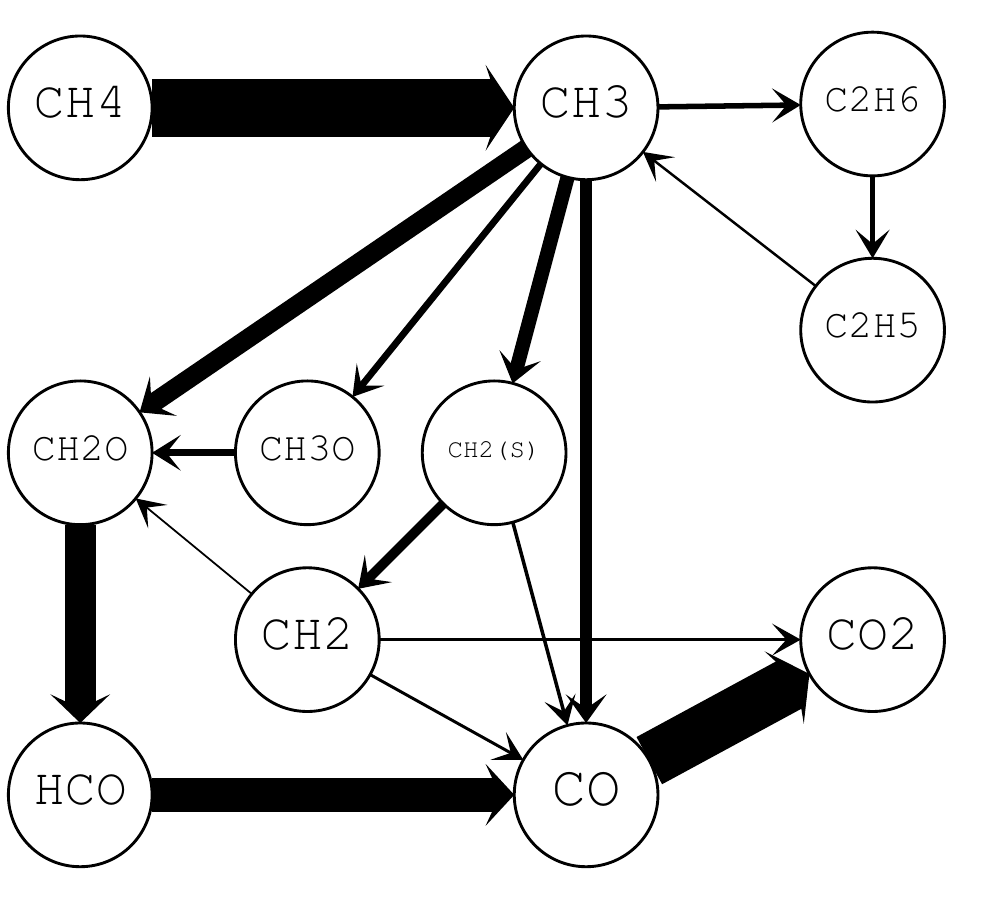}
\includegraphics[trim=140 175 150 125,clip,width=48mm]{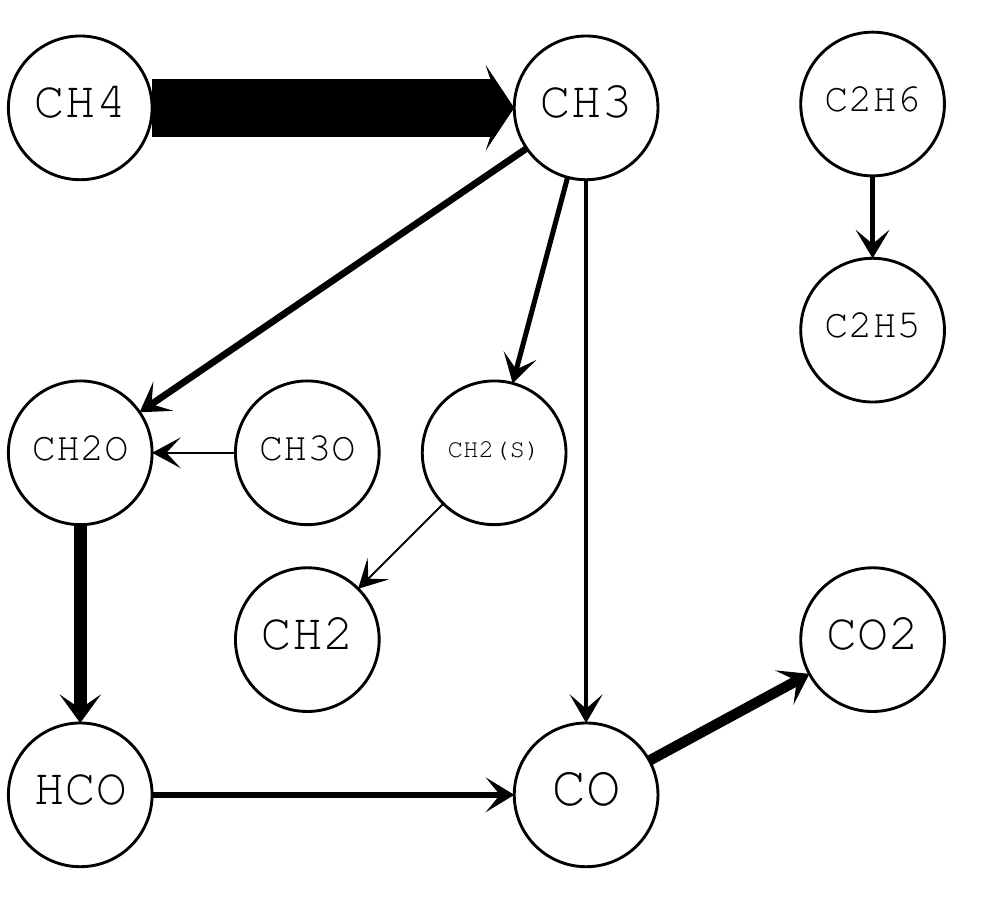}\\
\includegraphics[trim=120 200 120 70,clip,width=48mm]{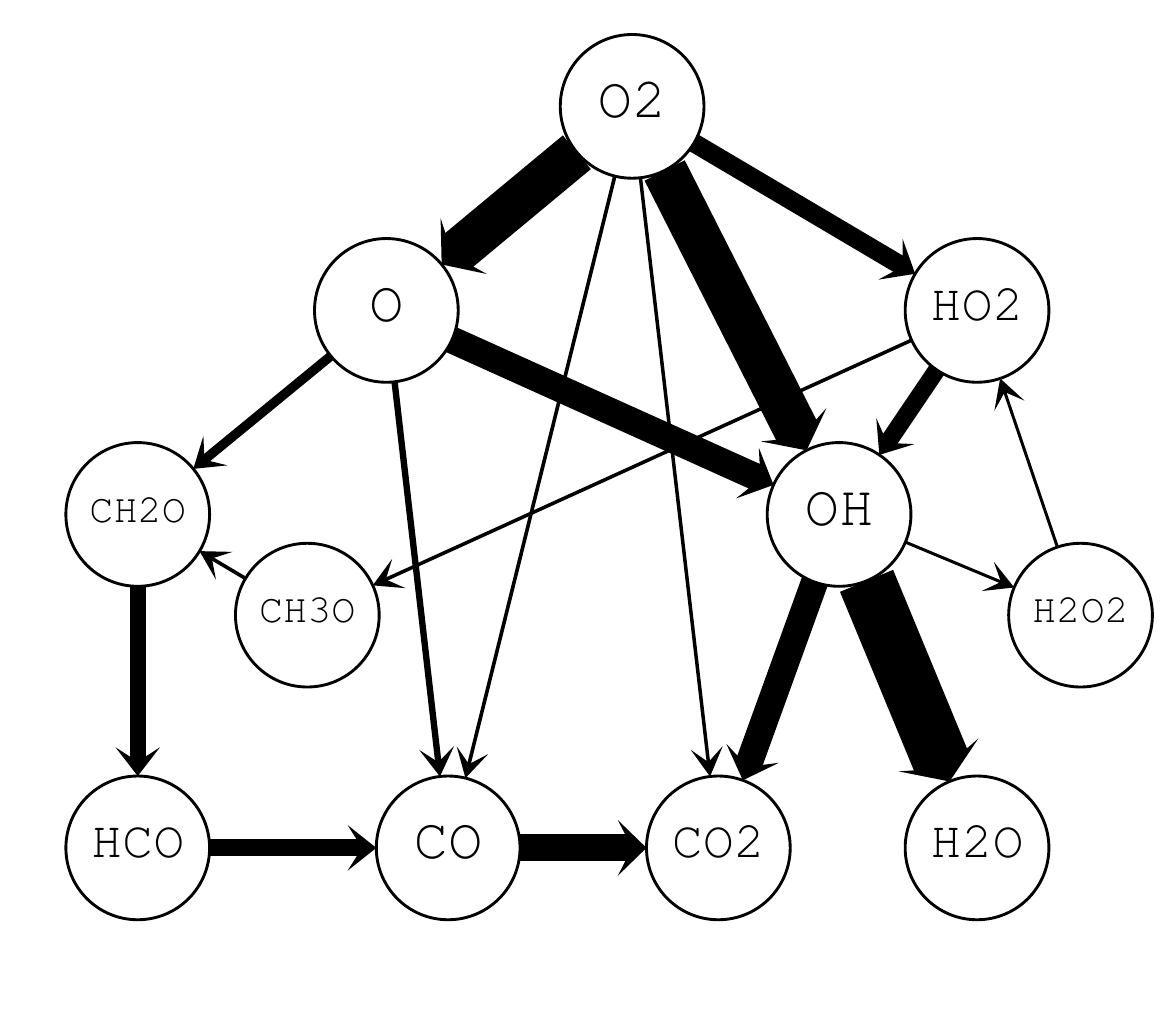}
\includegraphics[trim=120 200 120 70,clip,width=48mm]{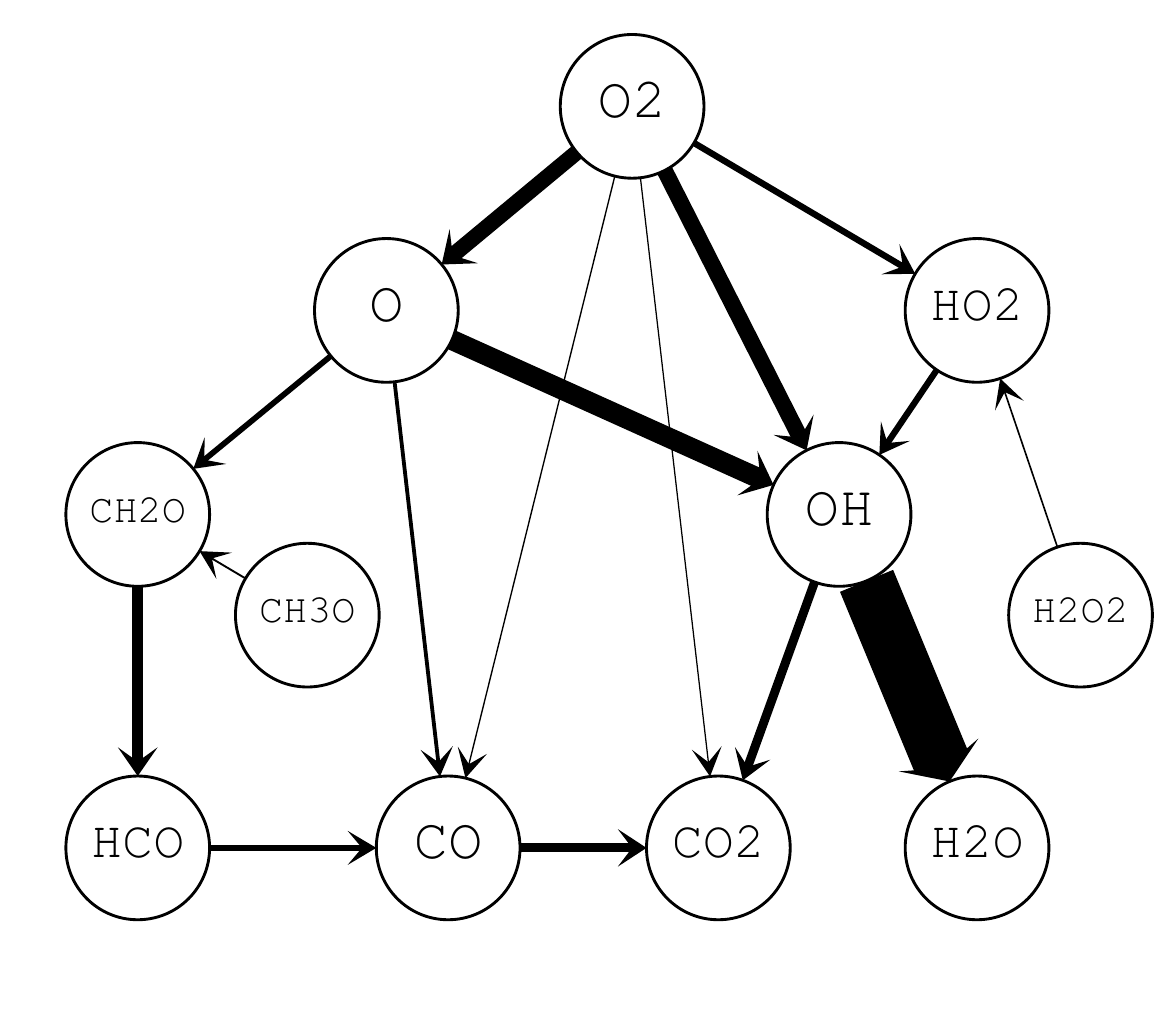}%\\
\caption{Reaction path diagrams following carbon (top) and oxygen (bottom);
unfiltered DNS data on the left, and reaction rates based on filtered species/temperature
DNS data on the right.}
\label{fig:reacPaths}
\end{figure}

Before considering individual reactions, the effect of the filter on the overall reaction paths
are considered.  Path diagrams for $\Ka=108$ are shown in figure~\ref{fig:reacPaths} for carbon 
(top), and oxygen (bottom), with the unfiltered DNS data on the left, and 
reaction rates based on filtered species/temperature DNS data on the right (note that the
reaction path diagram of filtered reactions is not presented as the operation used to
construct the reaction paths integrates out the effect of the filter).  
The size of each arrow reflects the rate of atom transfer normalised by the peak rate, and only 
rates greater than 2\% of the peak are shown.
(Note that missing links do not indicate that the reactions are not present, they have just 
fallen below the cut-off threshold based on the peak reaction rate.)
It is clear that the filter changes the balance of reaction paths
significantly; following carbon, the main decomposition of CH$_4$ to CH$_3$ far exceeds
all of the other rates, and following oxygen the final step of OH to H$_2$O becomes dominant.
The same response was observed for all Karlovitz numbers (see supplementary material, which
also includes path diagrams following hydrogen); 
the effect of the filter on the reaction rates
appears to be largely independent from Karlovitz number for these conditions.

\begin{figure}
\centering
\includegraphics[width=48mm]{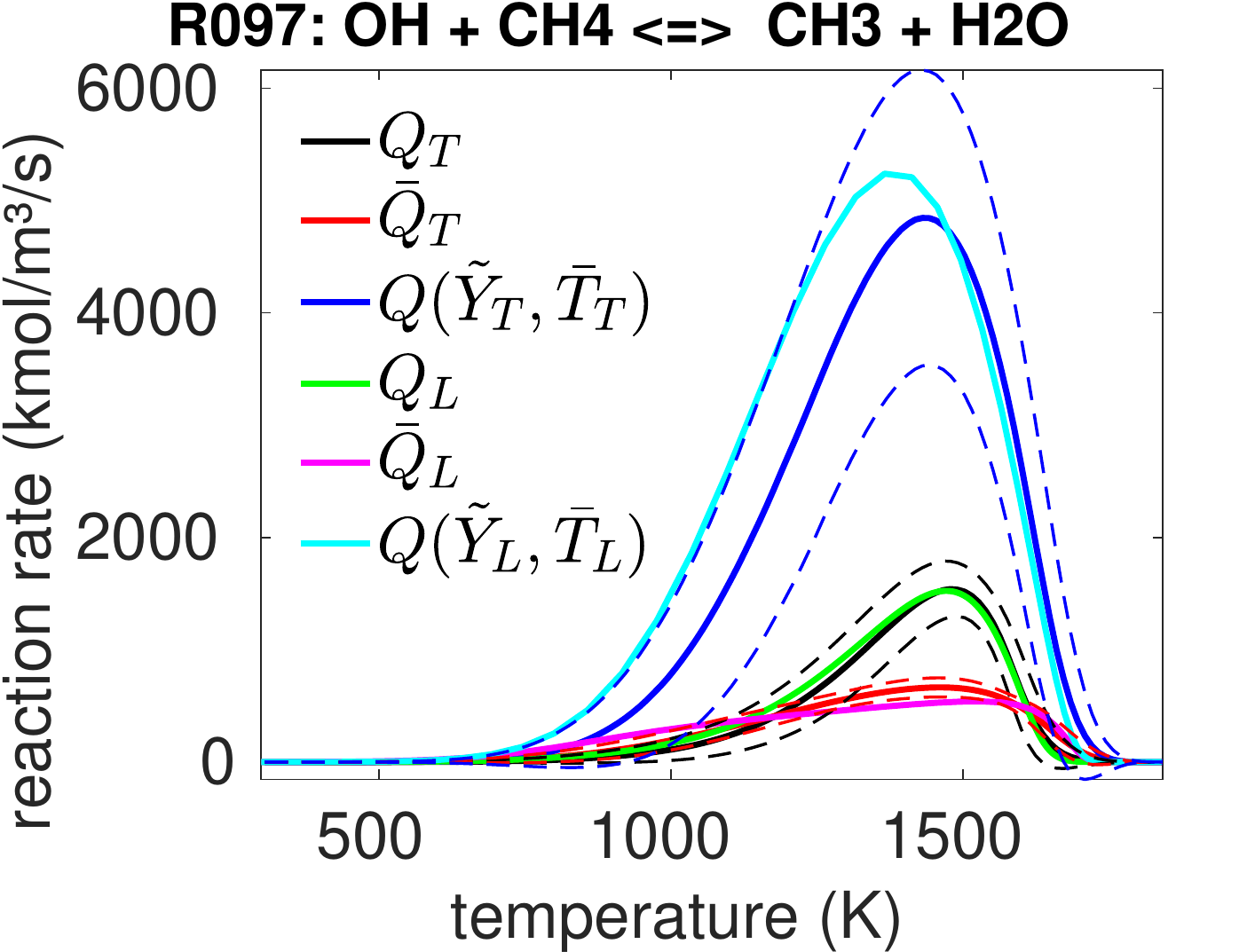}
\includegraphics[width=48mm]{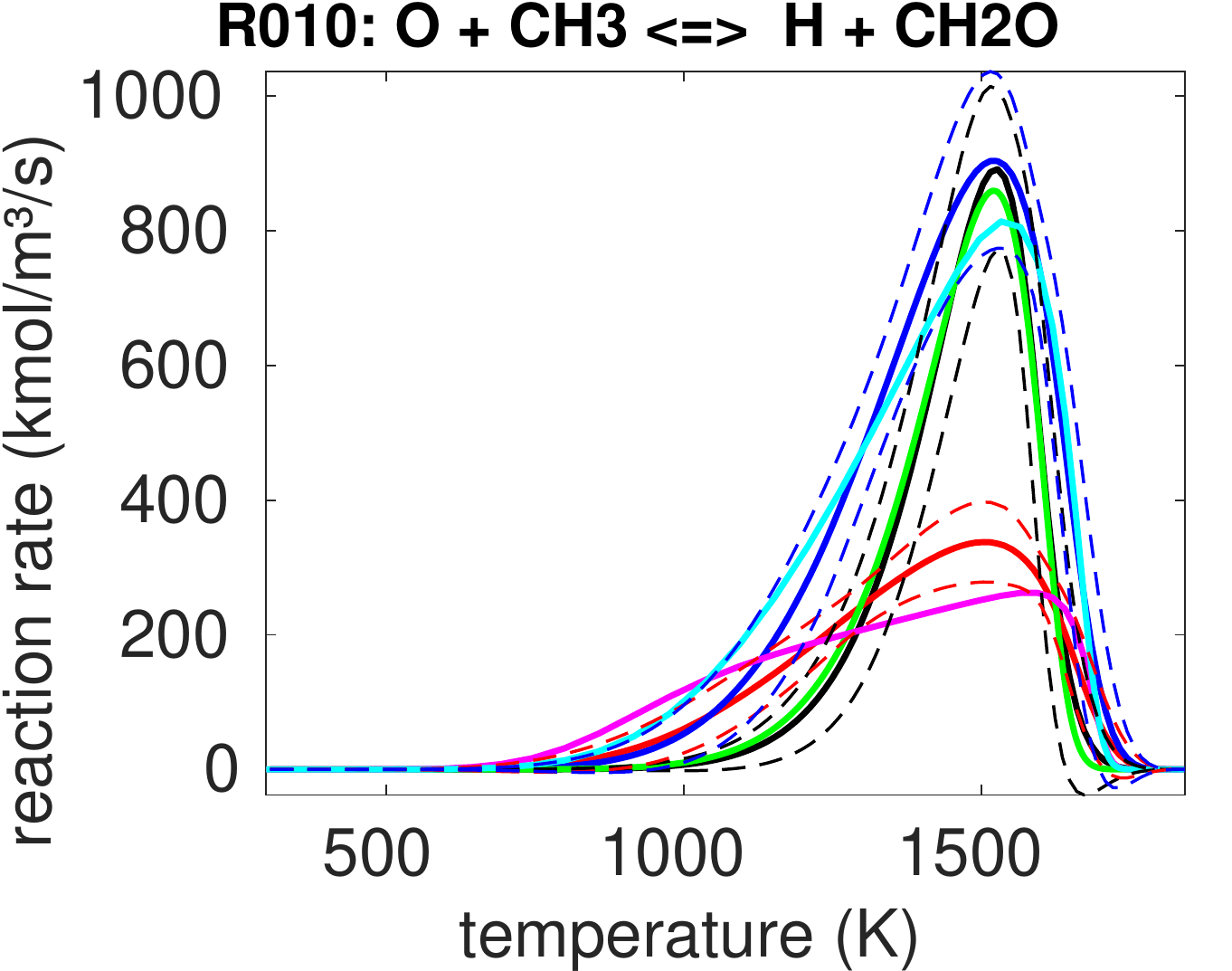}\vspace{2mm}\\
\includegraphics[width=48mm]{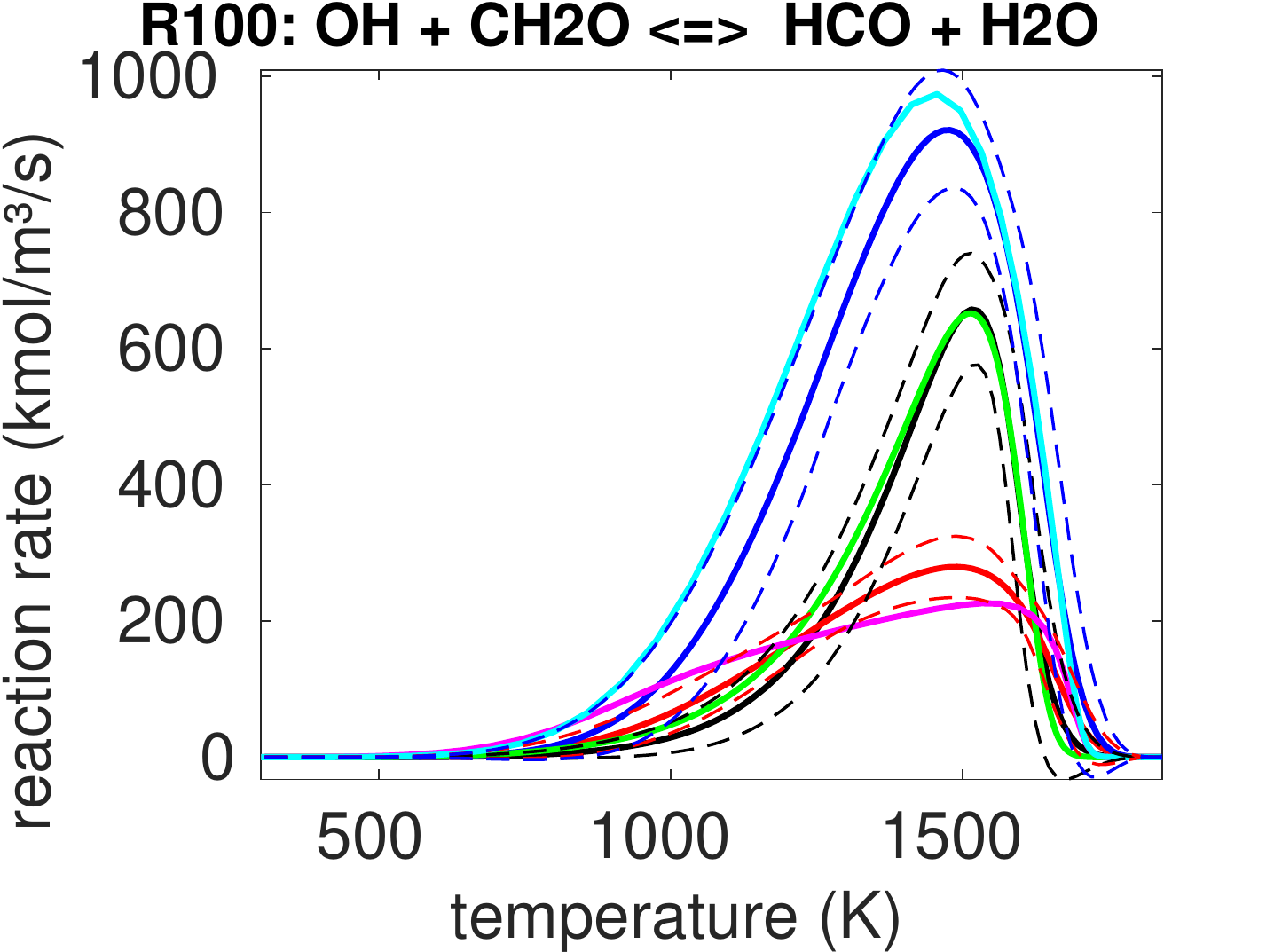}
\includegraphics[width=48mm]{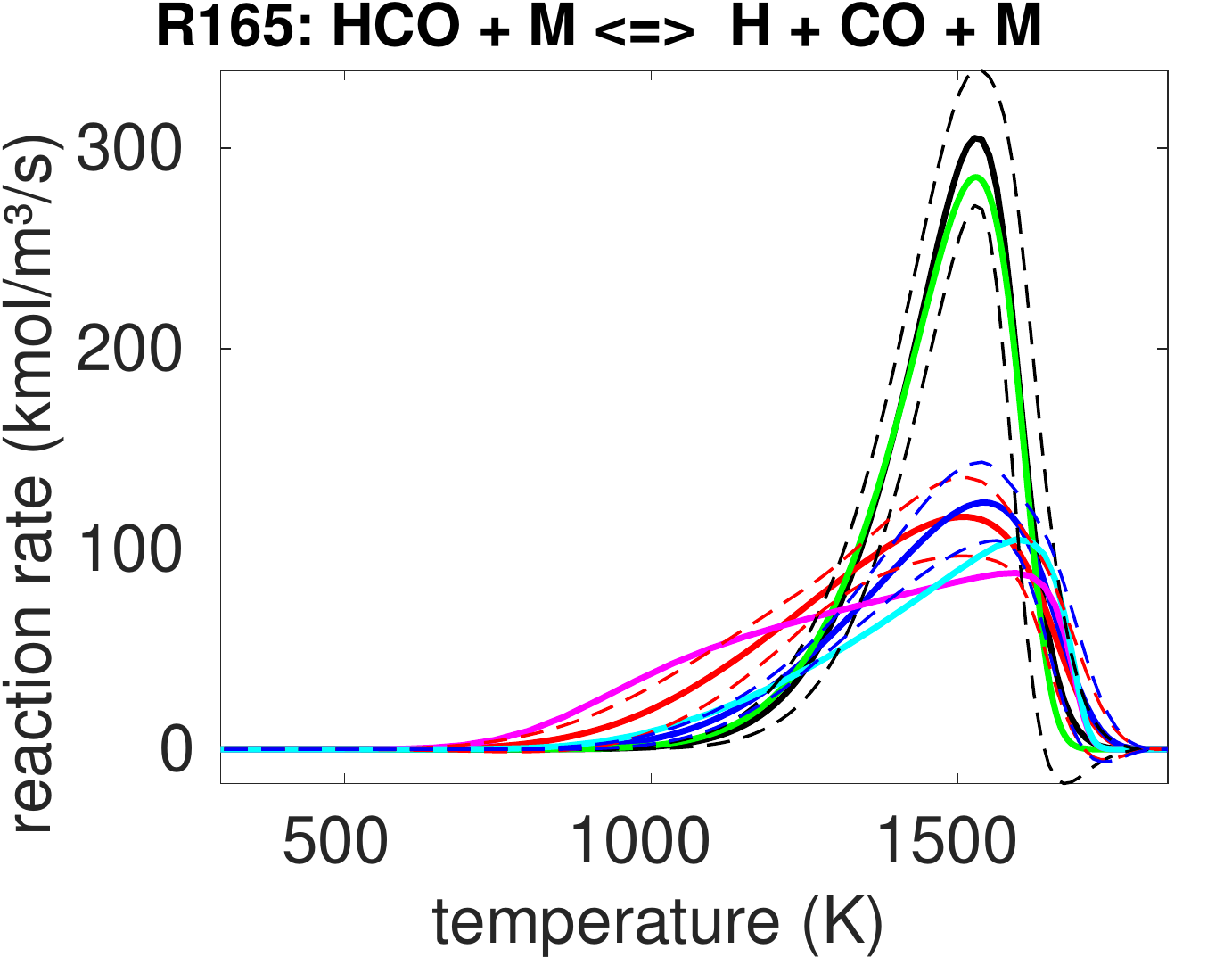}\vspace{2mm}\\
\includegraphics[width=48mm]{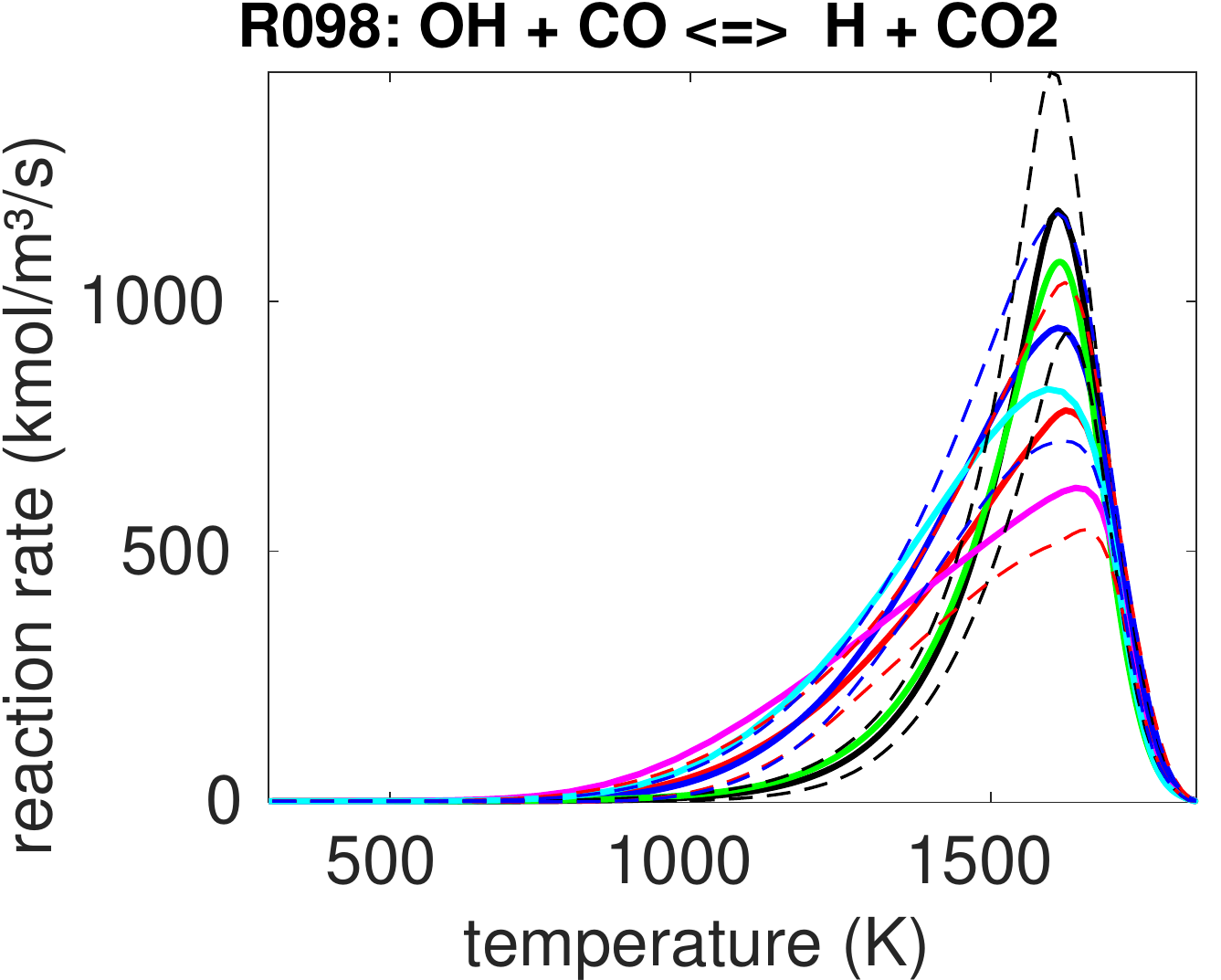}
\includegraphics[width=48mm]{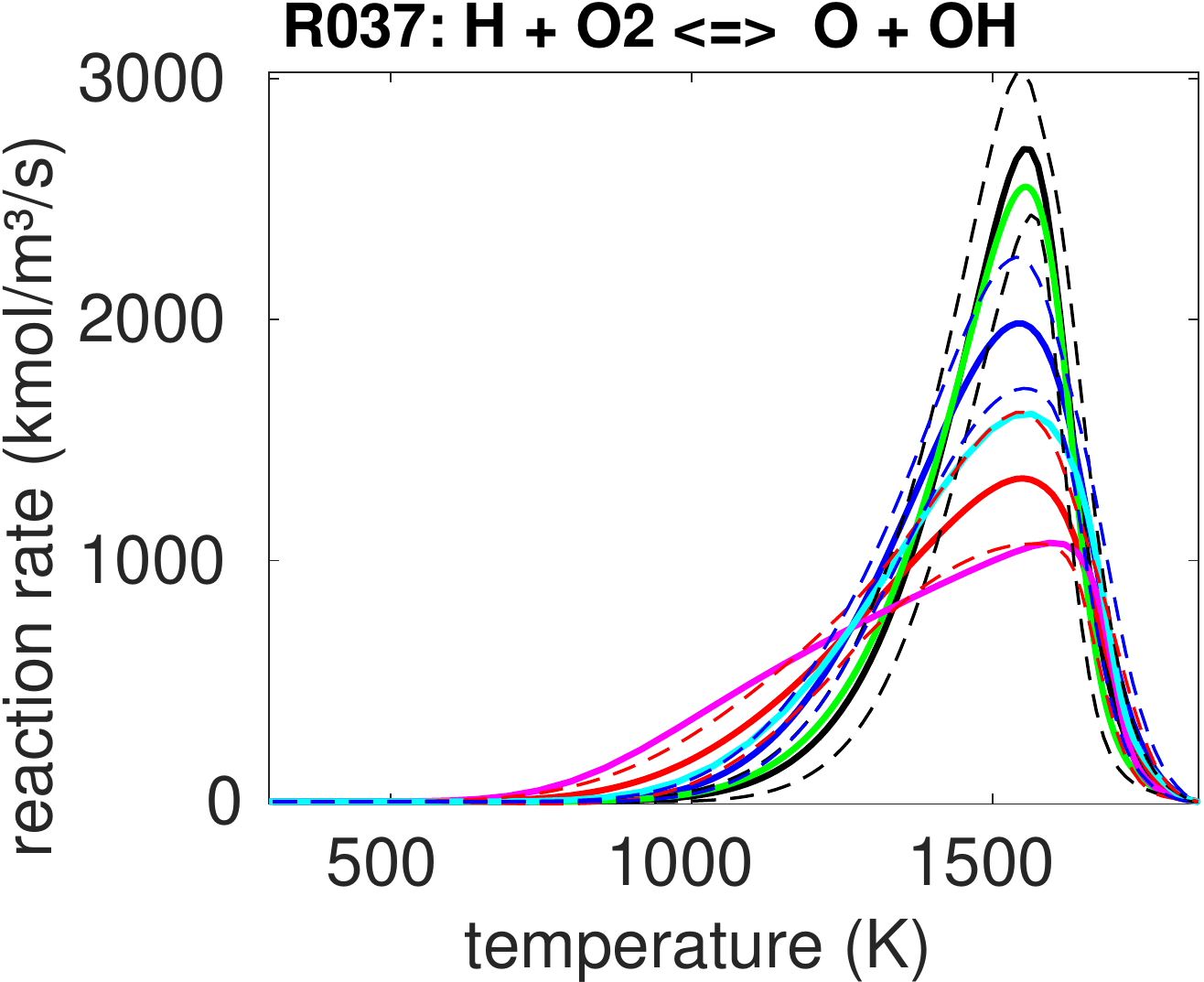}\vspace{2mm}\\
\includegraphics[width=48mm]{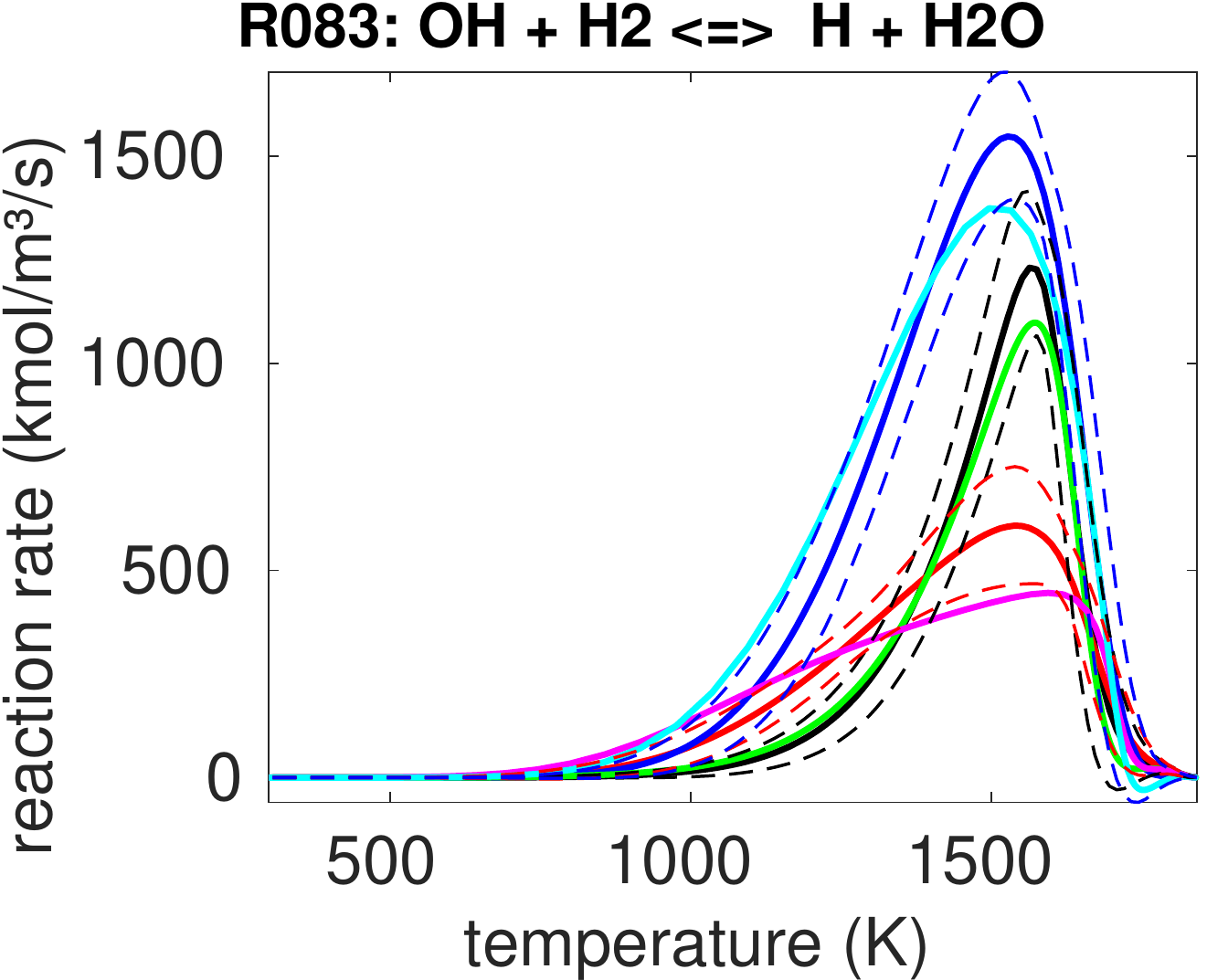}
\includegraphics[width=48mm]{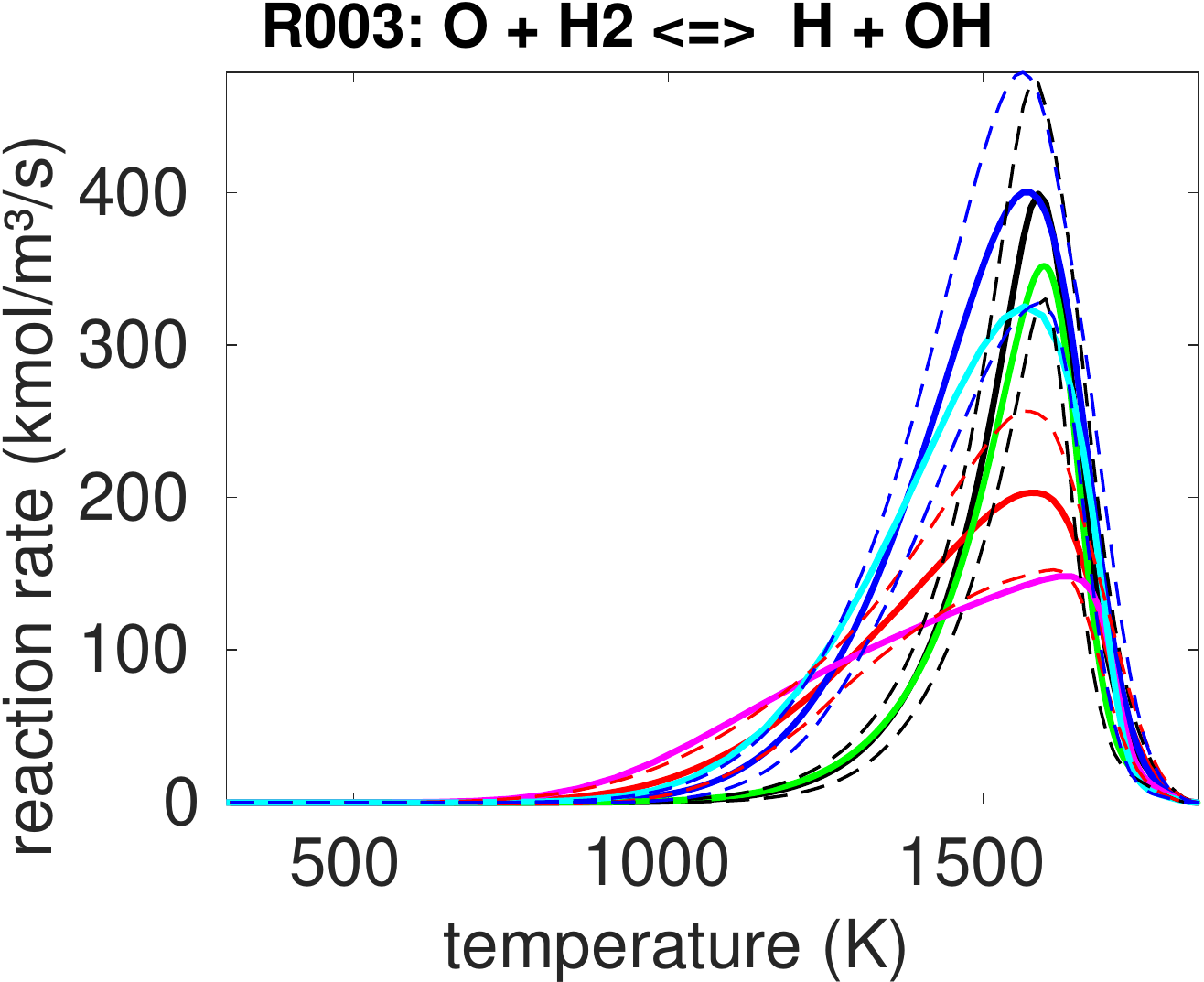}
\caption{Conditional means of filtered and unfiltered reaction rates for $\Ka=108$ and laminar flames.
Note that the filtered profiles are conditioned on filtered temperature.}
\label{fig:reactions}
\end{figure}

Profiles of the key reaction rates are presented in figure~\ref{fig:reactions} based on the
main pathway from the reaction path diagrams
(CH$_4$$\rightarrow$CH$_3$$\rightarrow$CH$_2$O$\rightarrow$HCO$\rightarrow$CO$\rightarrow$CO$_2$);
again, all reactions for all Karlovitz numbers are included as 
supplementary material for completeness.  In each plot, 
six reaction rate profiles are presented 
for unfiltered rates $Q_L$ and $Q_T$, filtered rates $\bar{Q}_L$ and $\bar{Q}_T$,
and rates evaluated with filtered species/temperature $Q(\tilde{\boldsymbol{Y}}_L,\bar{T}_L)$
and $Q(\tilde{\boldsymbol{Y}}_T,\bar{T}_T)$, where suffices $L$ and $T$ denote laminar and 
turbulent profiles, respectively; standard deviations are shown by dashed lines for turbulent cases.

The main reactions responsible for fuel consumption are hydrogen abstraction by O, H, and OH, the 
latter is shown by R097 in figure~\ref{fig:reactions}, but is representative of the other two 
reactions (see R011 and R052 in the supplementary material).  The conditional mean of the turbulent 
profile (black) aligns closely with the laminar profile (green), with similar alignment between the
filtered profiles (red and magenta).  The reaction rates of the filtered species/temperature (blue
and cyan) are significantly different from the other two pairs of profiles; the effects of 
non-linearities in reaction rates profoundly affects all three abstraction reactions, and explains 
the dominance of the CH$_4$$\rightarrow$CH$_3$ step in the carbon (and hydrogen) path diagrams of the 
filtered data.

The main reaction in the next step in the carbon path (CH$_3$$\rightarrow$CH$_2$O) is R010.
Once again, the reactions align closely in pairs, and while the reaction rates of filtered
species/temperature are high, they are not as significantly different as in the abstraction reactions.
The CH$_2$O$\rightarrow$HCO step is represented here by R100 (see also R015 and R057 in the
supplementary material), and shows an increase in reaction rates of filtered species/temperature with
a magnitude between that observed in R100 and the abstraction reations.
The HCO$\rightarrow$CO step (R165 presented here; see also R164 and R166 in the supplementary 
material) does not see a significant increase in magnitude of reaction rate, nor does the main
oxidation step CO$\rightarrow$CO$_2$ (R098).

The key reactions in the oxygen path diagram are R037, which takes O$_2$ to both O and OH, 
R083, which takes OH to H$_2$O (and H$_2$ to H), and R003 which take O to OH (and again H$_2$ to H); 
the latter stages of carbon oxidation are discussed above (R100, R165 and R098).
In R037, the effects of non-linearities are less pronounced than R083 (or the early stages in the 
carbon path), which explains the shift observed in the oxygen path diagram towards R083.
Again the reaction rate of filtered species/temperature (blue) aligns more closely with 
the corresponding laminar profile (cyan) than the turbulent filtered reaction profile (red).

For the reactions in general, it appears that in many cases, but by no means all, the profiles 
loosely align in pairs; specifically, the conditional mean of the turbulent data (black) aligns 
with the laminar profile (green), the filtered turbulent data (red) aligns with the filtered 
laminar profile (magenta), and the reaction rate of filtered turbulent species/temperature (blue) 
aligns with the reaction rates of the filtered laminar profile (cyan).
There are other reaction rates with interesting response to filtering (for example, see R033, R114 
and R124 in the supplementary material), especially those involving HO$_2$, H$_2$O$_2$ or CH$_3$O, 
but these are less significant reactions in the overall mechanism.

\begin{figure}
\centering
\includegraphics[width=48mm]{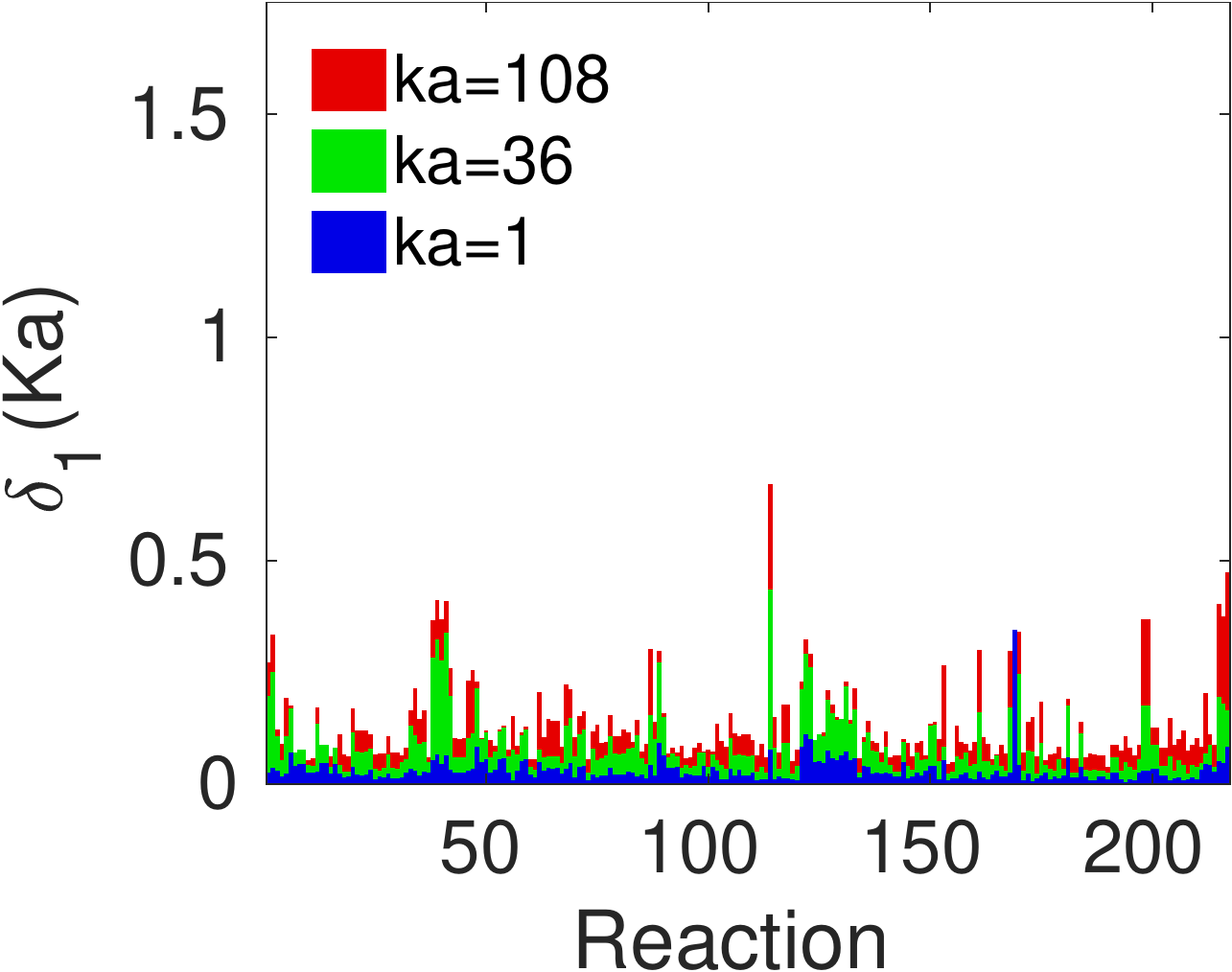}
\includegraphics[width=48mm]{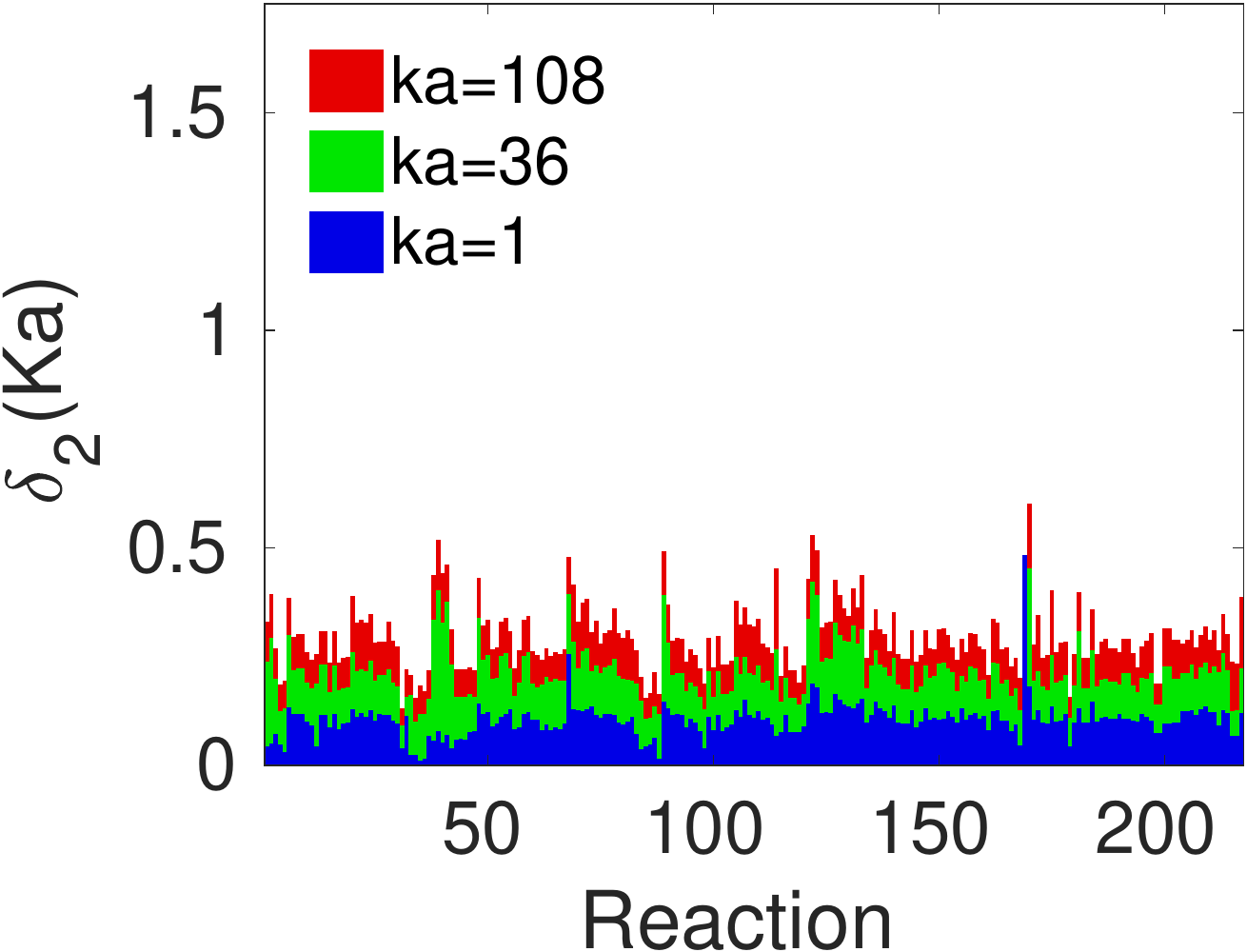}\vspace{2mm}\\
\includegraphics[width=48mm]{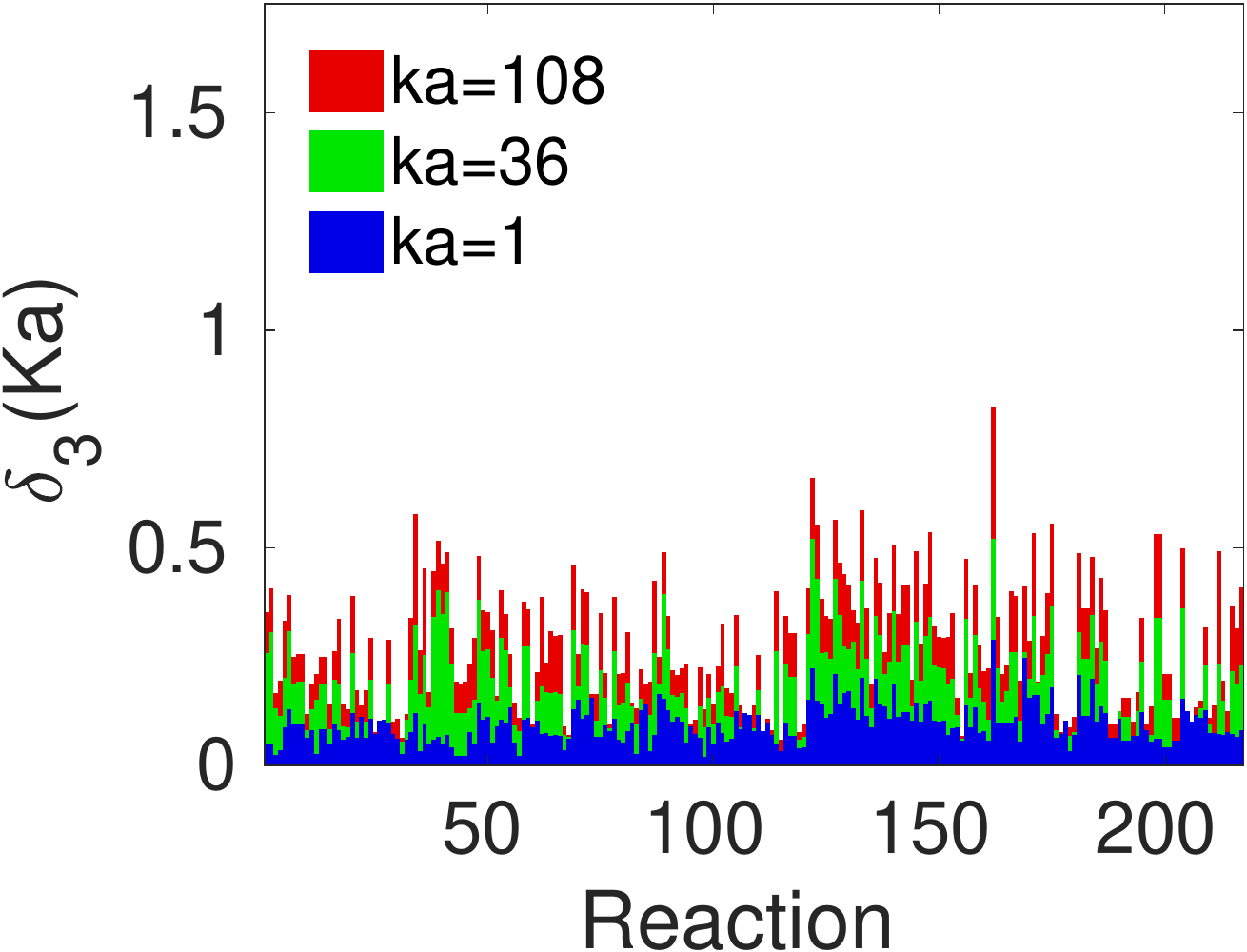}
\includegraphics[width=48mm]{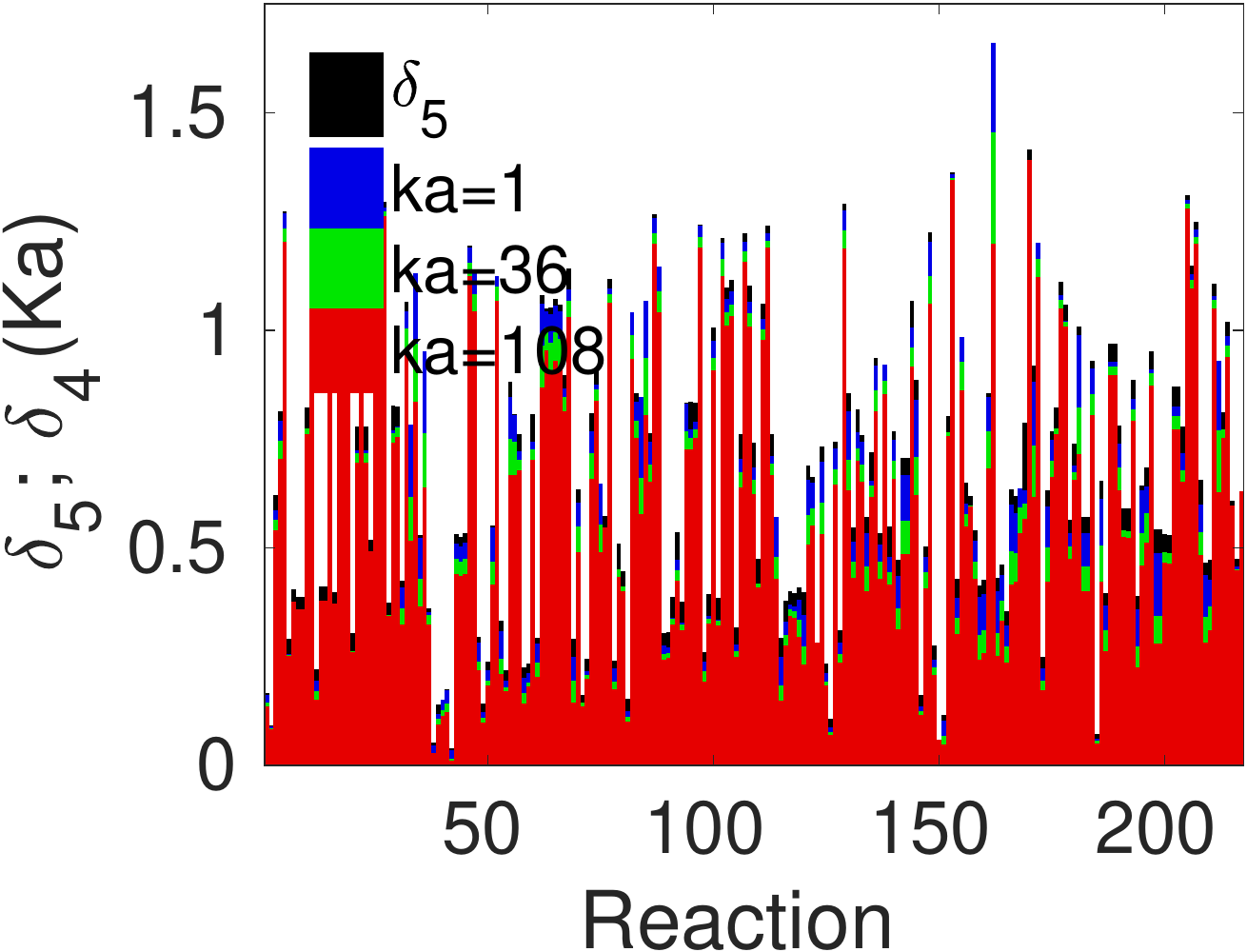}
\caption{Difference metrics $\delta_i$ for all turbulent cases.}
\label{fig:deltas}
\end{figure}

To quantify the alignment of the different reaction profiles, first define a normalised
difference between two functions $f(T)$ and $g(T)$ as
$$
\delta(f,g)=\sqrt{\int (f-g)^2\,\D T\Bigg/\frac{1}{2}\int f^2+g^2\,\D T},
$$
and then define five differences
\begin{equation*}
\delta_1=\delta\left(Q_T,Q_L\right),\qquad
\delta_2=\delta\left(\bar{Q}_T,\bar{Q}_L\right),
\end{equation*}
\begin{equation*}
\delta_3=\delta\left(Q(\tilde{Y}_T,\bar{T}_T),Q(\tilde{Y}_L,\bar{T}_L)\right),
\end{equation*}
\begin{equation*}
\delta_4=\delta\left(Q(\tilde{Y}_T,\bar{T}_T),\bar{Q}_T\right),\qquad
\delta_5=\delta\left(Q(\tilde{Y}_L,\bar{T}_L),\bar{Q}_L\right).
\end{equation*}
These five differences are depicted by bar graphs in figure~\ref{fig:deltas} for each Karlovitz 
number.  This comparison clearly demonstrates that the laminar-turbulent pairs of profiles from
figures~\ref{fig:species} and \ref{fig:reactions} (black and green by $\delta_1$, red and magenta by
$\delta_2$, and blue and cyan by $\delta_3$) are more closely aligned than the filtered reactions
(red, magenta) and the reaction of filtered species/temperature (blue, cyan) in both turbulent
and laminar differences ($\delta_4$, $\delta_5$).
Interestingly, the laminar difference $\delta_5$ is generally larger than all of the turbulent 
differences $\delta_4$, which actually decreases with increasing $\Ka$.

\begin{figure}
\centering
\includegraphics[width=48mm]{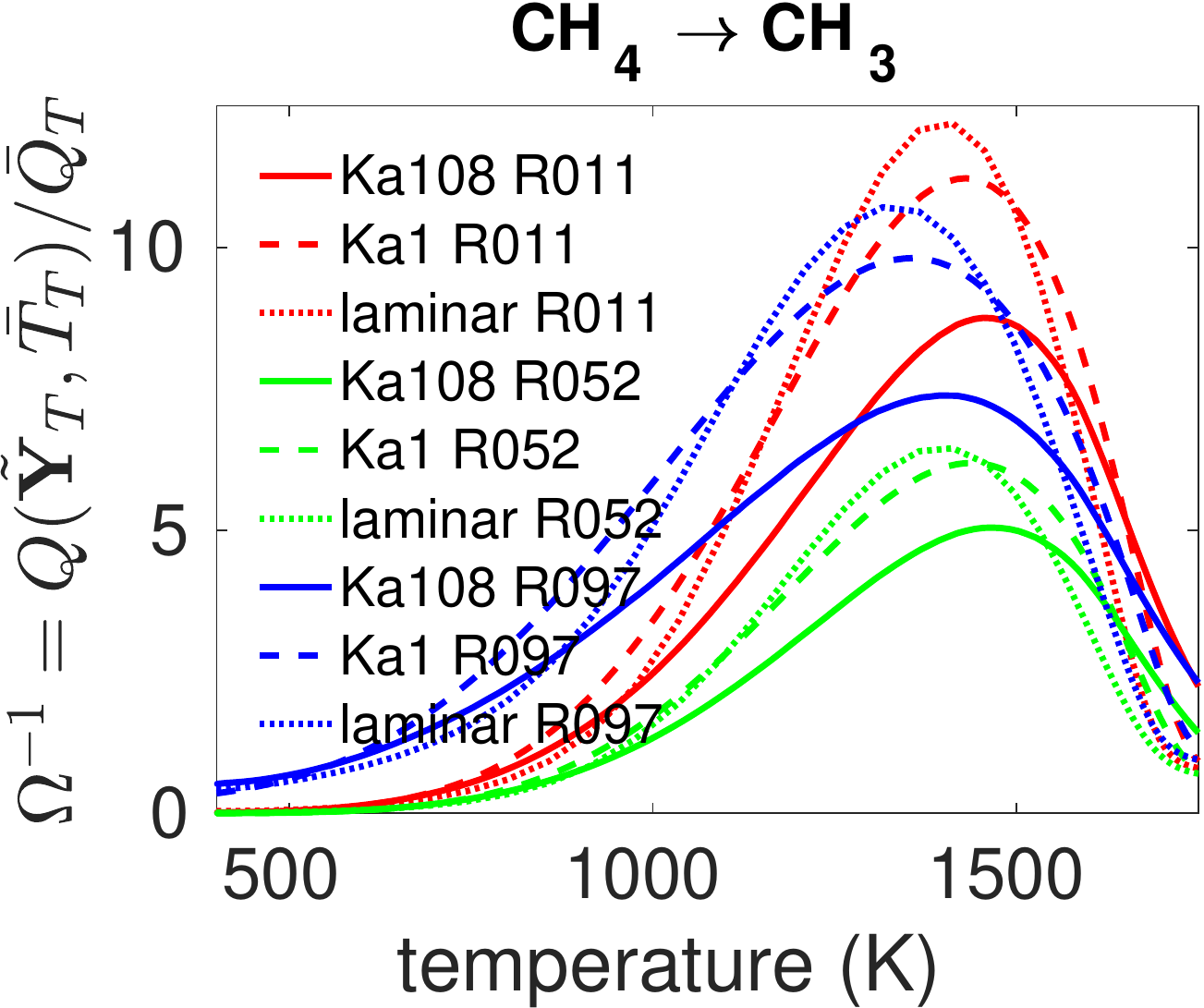}
\includegraphics[width=48mm]{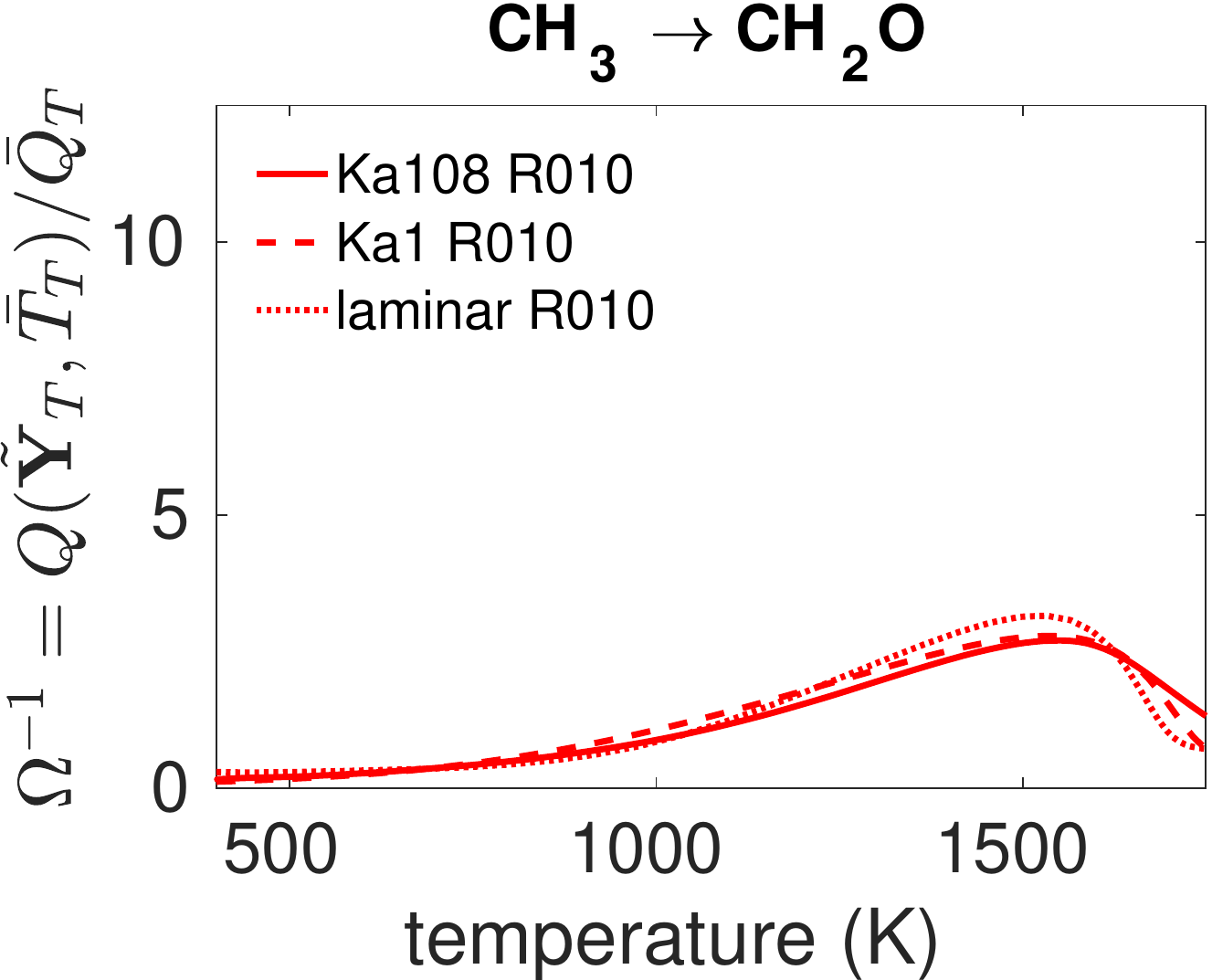}\vspace{2mm}\\
\includegraphics[width=48mm]{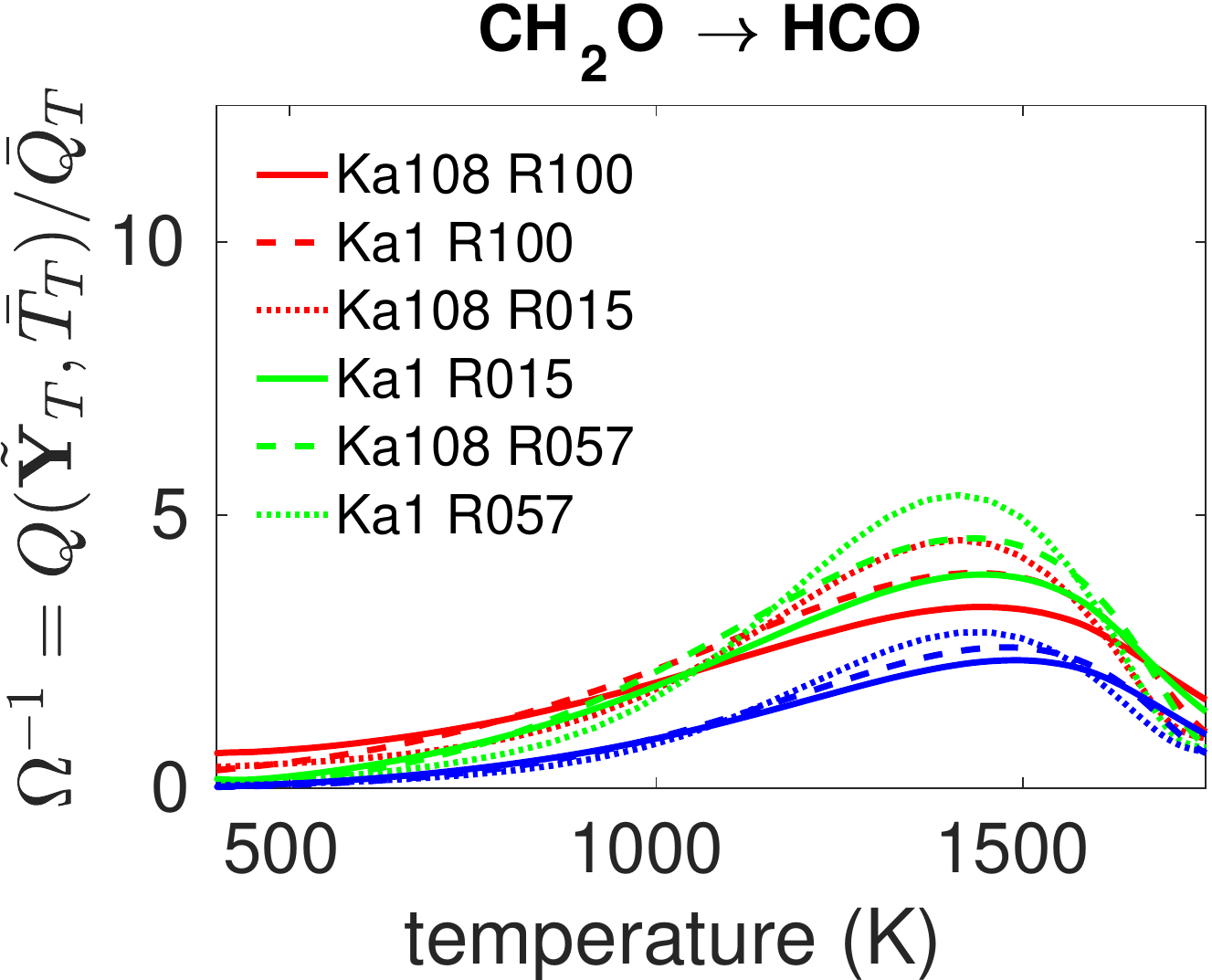}
\includegraphics[width=48mm]{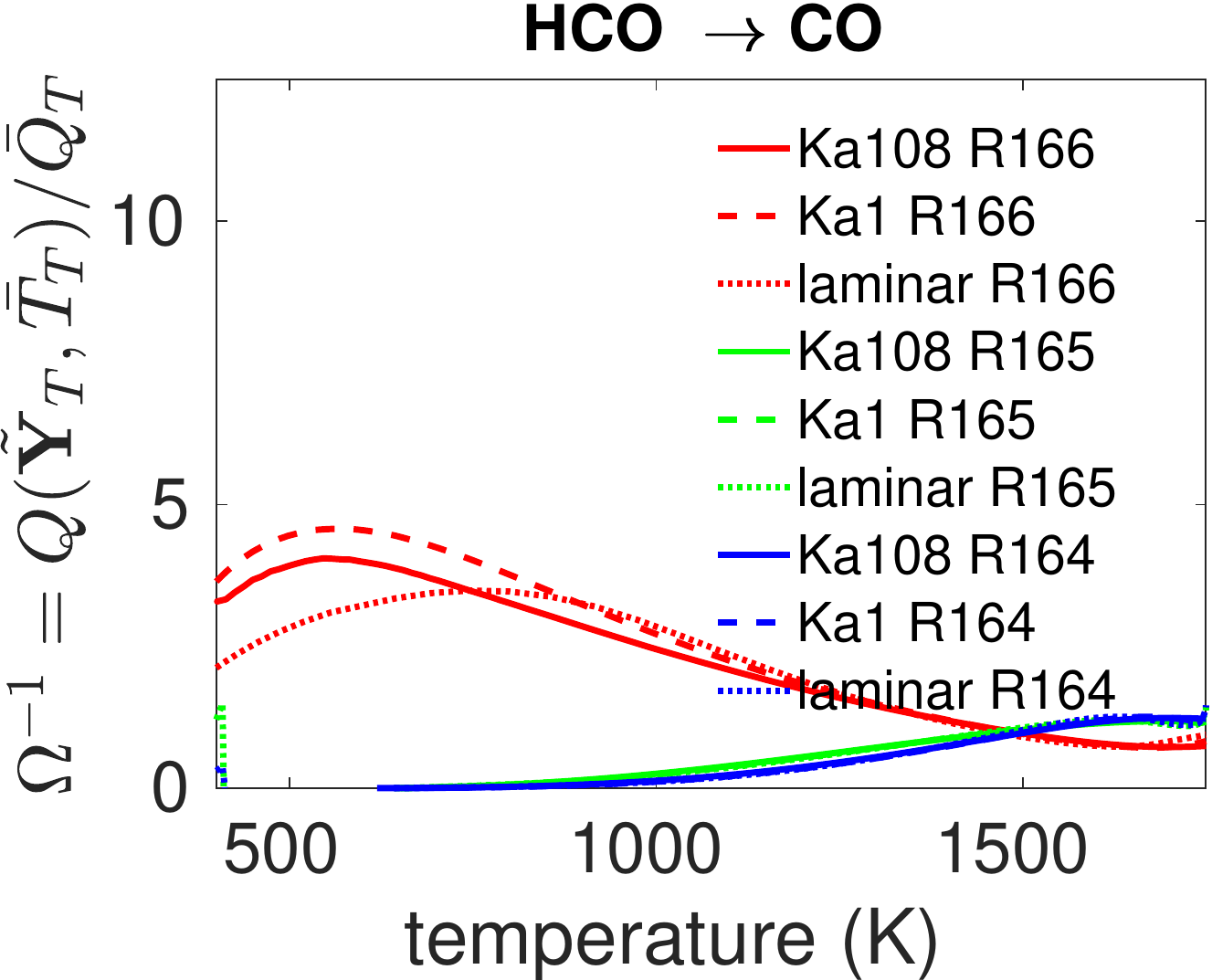}\vspace{2mm}\\
\includegraphics[width=48mm]{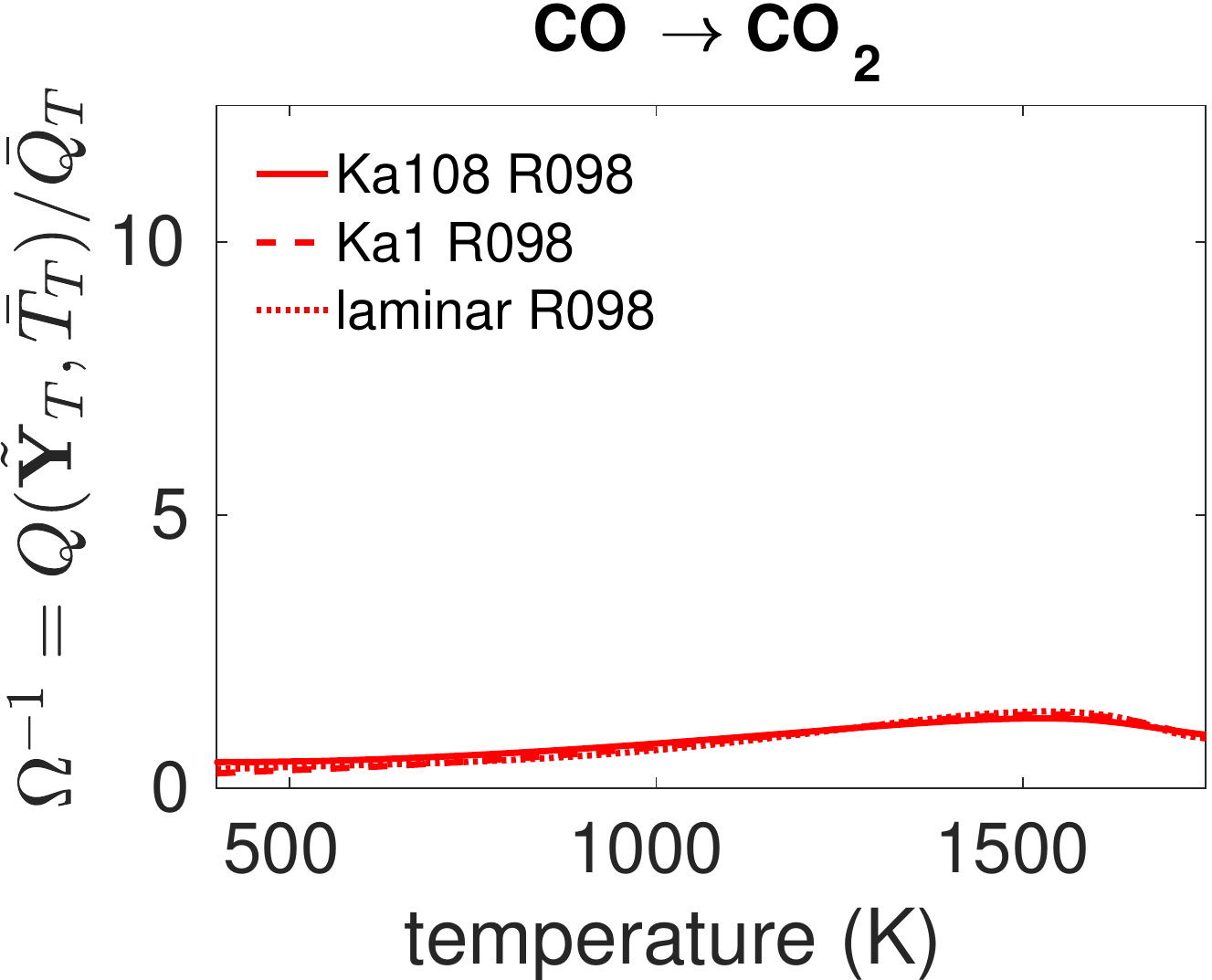}
\includegraphics[width=48mm]{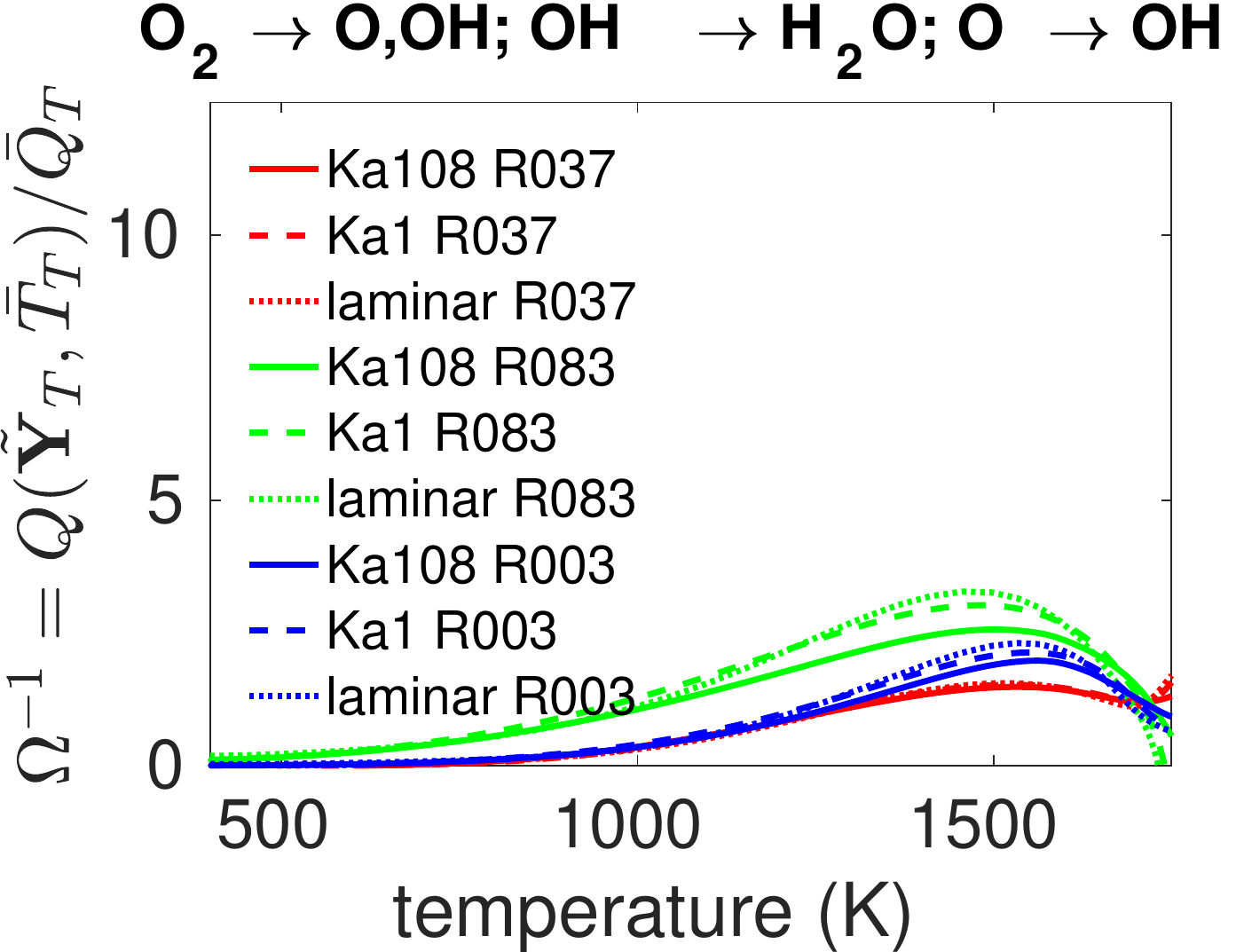}
\caption{Conditional means of reciprocals of reaction rate ratios, $\Omega_j^{-1}=Q_j(\tilde{\boldsymbol{Y}},\bar{T})/\bar{Q}_j$.}
\label{fig:omegas}
\end{figure}

From a modelling point of view, the ratio of the filtered reaction rate to the reaction rate of
filtered species/temperature was defined as
$\Omega_j=\bar{Q}_j/Q_j(\tilde{\boldsymbol{Y}},\bar{T})$.
Figure~\ref{fig:omegas} plots the reciprocal of $\Omega_j$ for the key reactions plotted in 
figure~\ref{fig:reactions}, along with other corresponding reactions in the key steps;
the reciprocal is used as it was found to tend to zero on both sides of the flame.
Once again, the strongest response is observed in the hydrogen abstraction reactions R011,
R052 and R097.
Perhaps surprisingly, there appears to be relative insensitivity to turbulence intensity;
furthermore, the profiles from the filtered laminar flame have also been plotted as dotted 
lines, and are reasonably close to the turbulent profiles.
This suggests that a potential turbulence modelling approach is to derived values for $\Omega_j$
based on the laminar flame.

\begin{figure}
\centering
\includegraphics[width=48mm]{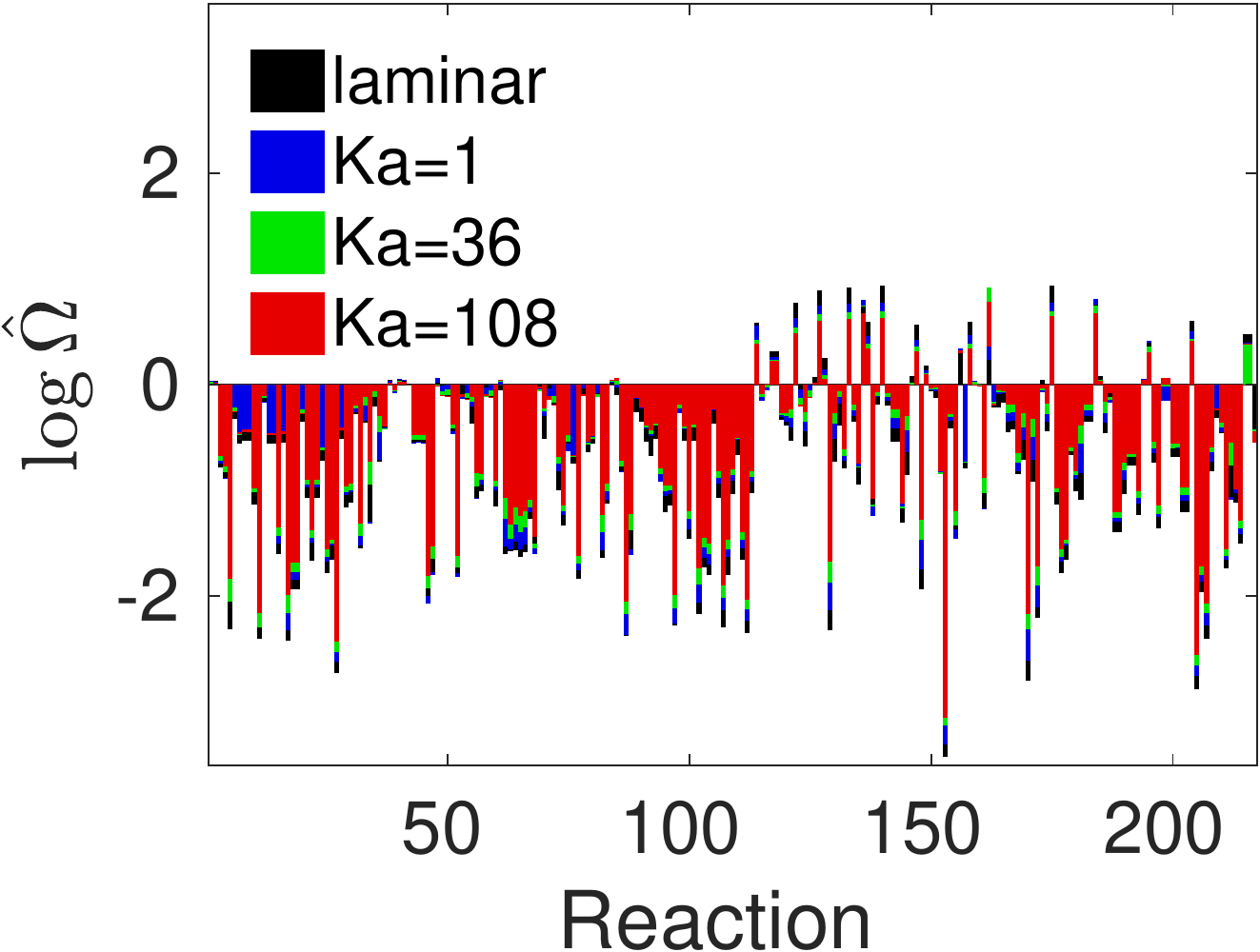}
\includegraphics[width=48mm]{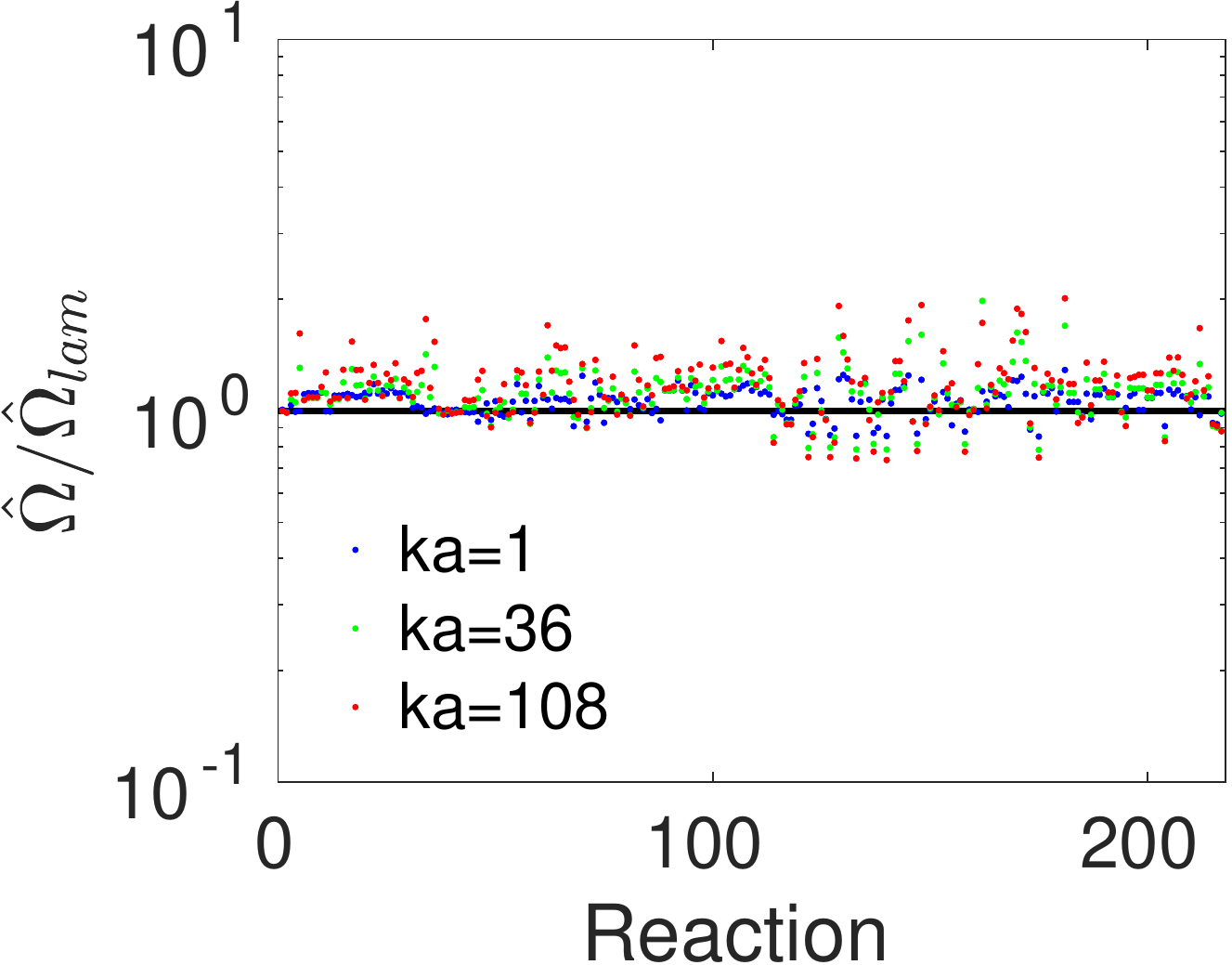}
\caption{Left: model values of $\log\hat{\Omega}_j$.  Right: $\hat{\Omega}_{j,T}/\hat{\Omega}_{j,L}$.}
\label{fig:modelOmega}
\end{figure}

Notwithstanding the temperature dependence of $\Omega_j$, a model constant $\hat{\Omega}_j$ can be 
defined for each reaction as
\begin{equation}
\hat{\Omega}_j=\frac{\max_T\left|\bar{Q}_{j}\right|}{\max_T\left|Q_j(\tilde{\boldsymbol{Y}},\bar{T})\right|}.
\end{equation}
This represents a simple scaling of each reaction based on the ratio of filtered reaction rate to 
reaction rate of filtered species and temperature; note that the errors introduced through this
approximation will be greatest away from the flame where the reactions go to zero, and so are
expected to be of little importance.  The log of $\hat{\Omega}_j$ is presented in
figure~\ref{fig:modelOmega}, which gives equal weighting to reactions that need to be enhanced
(positive values) and those the need to be suppressed (negative values).  The values are 
predominantly negative, indicative that $Q_j(\tilde{\boldsymbol{Y}},\bar{T})$ is generally 
higher than $\bar{Q}_j$.  Pronounced negative values are found for the abstraction reactions 
considered above, along with those for ethane and propane (e.g.~R0027, R205 and R207), and some
other reactions involving CH$_3$O (e.g.~R017 and R153).  Pronounced positive values appear to involve
species and reactions that have narrow profiles, and so are smeared by the filter (e.g. R122, R127, 
R133, R140, and R175).  Again, note that there is relative insensitivity to $\Ka$, and typically, 
the highest $\Ka$ corresponds to the smaller values of $\log\hat{\Omega}_j$.
Scaling the turbulent $\hat{\Omega}_{j,T}$ by the corresponding laminar value, $\hat{\Omega}_{j,L}$, 
as shown in the right-hand panel of figure~\ref{fig:modelOmega}, further demonstrates a surprising 
insensitivity to turbulence conditions.

\begin{figure}
\centering
\includegraphics[trim=140 175 150 125,clip,width=48mm]{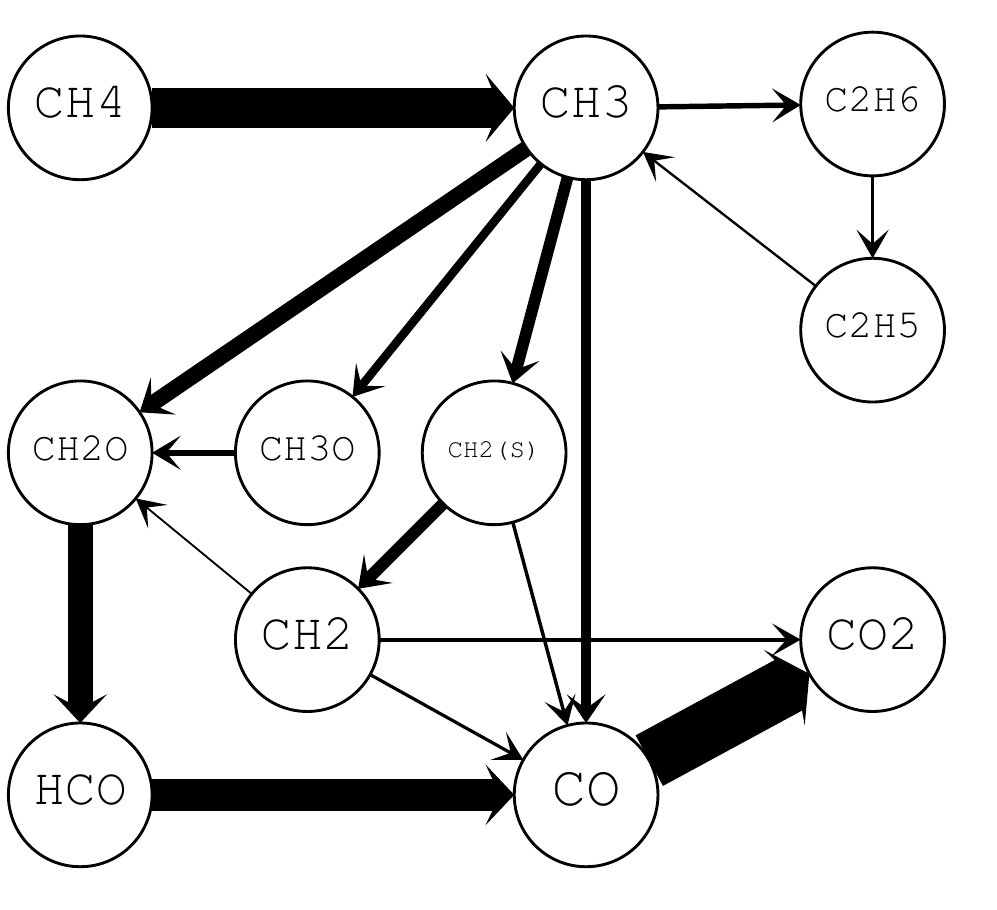}
\includegraphics[trim=120 200 120 70,clip,width=48mm]{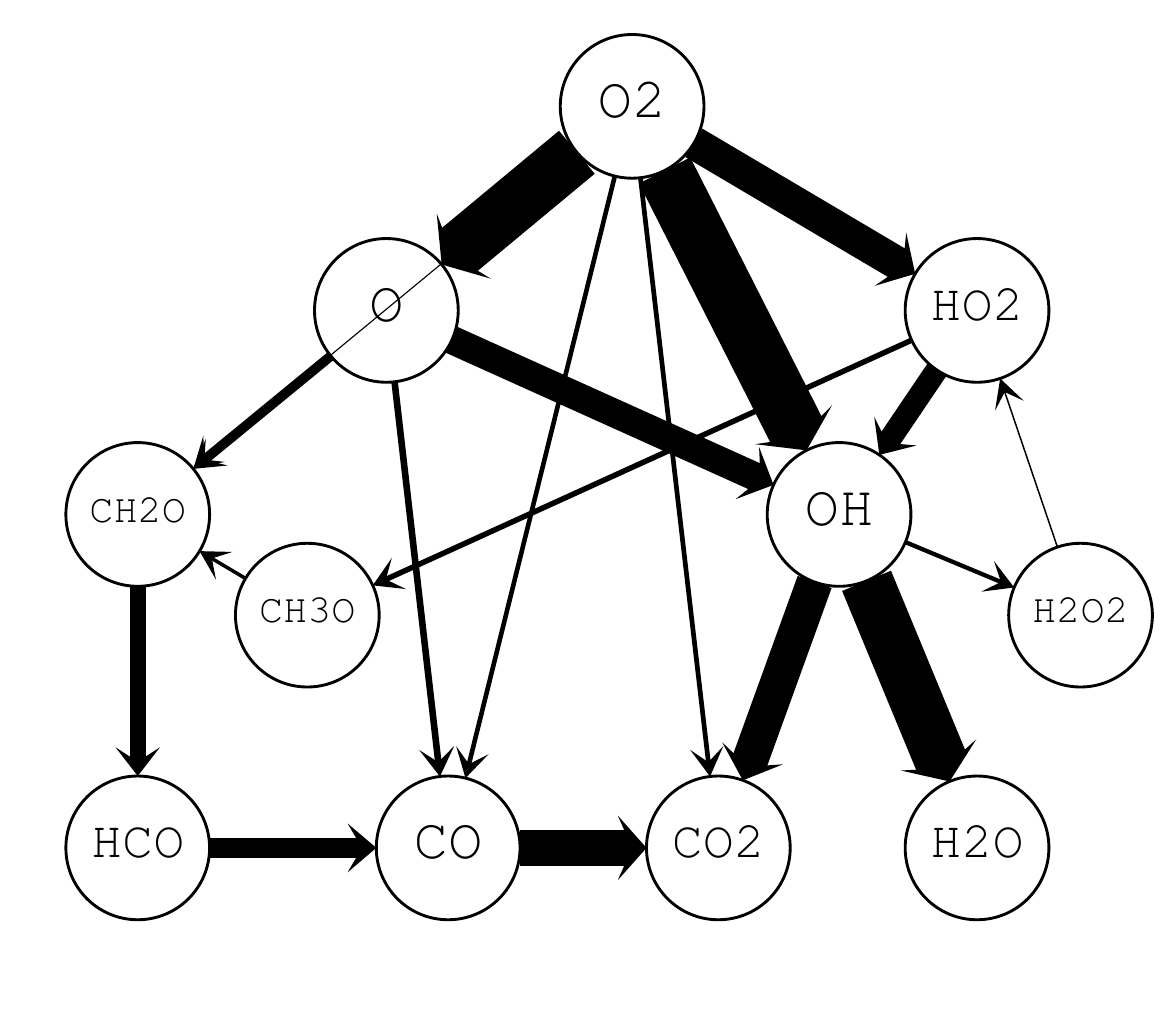}
\caption{Reaction path diagrams following carbon (left) and oxygen (right) using the model values of 
$\hat{\Omega}_{j,L}$ and filtered species and temperature from the $\Ka=108$ case.}
\label{fig:modelPaths}
\end{figure}

This suggests that it may be possible to formulate a model for $\hat{\Omega}_j$ based on filtered 
laminar profiles alone.  To examine this potential, reaction path diagrams are presented in 
figure~\ref{fig:modelPaths}
that use filtered species and temperature from the $\Ka=108$ case with reaction rates modified
using $\hat{\Omega}_{j,L}$ from the laminar flame profiles.  Compared with the corresponding plots on
the figure~\ref{fig:reacPaths} the modified reaction paths align much more closely with the 
filtered reaction rates than the reaction rates of the filtered species and temperature.  
There are some clear
differences; the CH$_4$$\rightarrow$CH$_3$ edge is thinner in figure~\ref{fig:modelPaths}, and 
there are some new reaction paths present in the oxygen path in figure~\ref{fig:modelPaths}
(e.g. O$_2$$\rightarrow$CH$_2$O and HO$_2$$\rightarrow$CH$_3$O).  These differences suggest that 
the balance has not been completely restored by the model $\hat{\Omega}_{j,L}$, and the normalisation
hides the overall reaction rates (which can be easily tuned), but this example provides 
proof-of-concept that a simple scaling of reaction rates by $\hat{\Omega}_{j,L}$ based on filtered 
laminar profiles is a straightforward approach that may yield feasible flame models for LES with 
finite-rate chemistry.

\section{Discussion and Conclusions}

An {\it a priori} analysis of a DNS database of turbulent lean premixed methane flames has been
presented.  The leading-order effect was found to be due to the 
filter operation, and flame response to turbulence was a secondary effect 
(figure~\ref{fig:deltas}), which manifested primarily as an increase in standard deviation; 
moreover, with increasing Karlovitz 
number, the disparity (as represented by $\delta_4$) was found to decrease.
Species profiles (figure~\ref{fig:species}) were found to align with the classification presented 
in \cite{AspdenCNF16}.  Importantly, the radicals O, H and OH were found
to be less impacted by the filter than other high-temperature radicals, which were significantly
reduced in magnitude by the filter.  It is the non-linear response in the reaction progress rates 
that presents the main modelling challenge (figure~\ref{fig:reactions}).
By considering reaction path diagrams, key reactions have been identified that are responsible for 
disparities between the desired filtered reaction rates and the reaction rates evaluated using 
quantites available in LES calculations (i.e.\ the filtered species and temperature).  
Specifically, the hydrogen abstraction reactions that take CH$_4$ to CH$_3$ (by O, H and OH)
were found to have a particularly enhanced reaction rate, and dominate the whole reaction path 
diagram.  Under the conditions presented, reaction paths were found to be largely 
independent of turbulence intensity.  
In general, reaction rates were found to align in pairs; i.e.~the turbulent profile $Q_T$
aligned with the laminar profile $Q_L$, the filtered profiles $\bar{Q}_T$ aligned with $\bar{Q}_L$,
and the reaction rate of filtered quantities $Q(\tilde{\boldsymbol{Y}}_T,\bar{T}_T)$ aligned with
$Q(\tilde{\boldsymbol{Y}}_L,\bar{T}_L)$, (see figure~\ref{fig:deltas}).
This alignment and relative insensitivity to $\Ka$ suggests that a model for the reaction rate 
scalings (e.g.~$\hat{\Omega}_{j,L}$) can be formulated based on filtered laminar profiles.  
To this end, an example
was considered by taking the ratio of the maximum absolute values, from which reaction paths were
formed that presented better agreement with the actual reaction paths than those evaluated using
filtered species and temperature.  The example given is not intended to be a model proposal, 
and much more work is required to develop this concept into a predictive model.
In particular,
further work is required to consider a broader range of conditions (e.g.\ other fuels, Lewis 
number effects, and turbulent conditions), and the effect of the filter width (and form).
Moreover, such {\it a priori} analysis is a long way from a predictive model that will perform 
well in practice; it is anticipated that the simple approach proposed here will perform poorly 
without further calibration.  Fine tuning the model constants for reaction rates could be performed
by some kind of automated approach.  Finally, confirmation of
the approach will have to come from successful {\it a posteriori} testing, but the present paper
demonstrates proof-of-concept of a potential approach for formulating a
reaction rate model for LES with finite-rate chemistry, and highlights the possibility of 
being able to base the approach on filtered laminar flames.

\section*{Acknowledgments}
\label{Acknowledgments}

CF and NZ acknowledge the financial support from the Swedish Armed forces and by the 
Swedish Energy Agency via the EFFECT2 project. The authors are also grateful to
John Bell and Marc Day for computational support.

\bibliography{paper} 
\bibliographystyle{unsrt}

\end{document}